\documentclass[aps,prl,twocolumn,superscriptaddress]{revtex4}

\usepackage{graphicx,amssymb,amsfonts,amsmath,chemarr,color,commath,braket}

\begin{document}

\title{Precision of flow sensing by self-communicating cells}

\author{Sean Fancher}
\affiliation{Department of Physics and Astronomy, Purdue University, West Lafayette, IN 47907, USA}
\affiliation{Department of Physics and Astronomy, University of Pennsylvania, Philadelphia, PA 19104, USA}

\author{Michael Vennettilli}
\affiliation{Department of Physics and Astronomy, Purdue University, West Lafayette, IN 47907, USA}

\author{Nicholas Hilgert}
\affiliation{Department of Physics and Astronomy, Purdue University, West Lafayette, IN 47907, USA}

\author{Andrew Mugler}
\email{amugler@purdue.edu}
\affiliation{Department of Physics and Astronomy, Purdue University, West Lafayette, IN 47907, USA}

\begin{abstract}
Metastatic cancer cells detect the direction of lymphatic flow by self-communication: they secrete and detect a chemical which, due to the flow, returns to the cell surface anisotropically. The secretion rate is low, meaning detection noise may play an important role, but the sensory precision of this mechanism has not been explored. Here we derive the precision of flow sensing for two ubiquitous detection methods: absorption vs.\ reversible binding to surface receptors. We find that binding is more precise due to the fact that absorption distorts the signal that the cell aims to detect. Comparing to experiments, our results suggest that the cancer cells operate remarkably close to the physical detection limit. Our prediction that cells should bind the chemical reversibly, not absorb it, is supported by endocytosis data for this ligand-receptor pair.
\end{abstract}

\maketitle

Metastasis is the process of cancer cells spreading from the primary tumor to other parts of the body. A major route for spreading is the lymphatic system, a network of vessels that carry fluid to the heart. Particular cancer cells detect the drainage of lymphatic fluid toward the vessels and move in that direction \cite{nathanson2003insights}. Experiments have shown that the detection occurs by self-communication: the cells secrete diffusible molecules (CCL19 and CCL21) that they detect with receptors (CCR7) on their surface \cite{shields2007autologous}. The flow affects the distribution of detected molecules thereby provides information about the flow direction. This flow detection mechanism, termed `autologous chemotaxis,' has been observed for breast cancer \cite{shields2007autologous}, melanoma \cite{shields2007autologous}, and glioma cell lines \cite{munson2013interstitial}, as well as endothelial cells \cite{helm2005synergy}, and has been studied using fluid dynamics models \cite{shields2007autologous, fleury2006autologous, waldeland2018multiphase}.

The flow is slow. Lymphatic drainage speeds near tumors are typically $v_0 = 0.1$$-$$1$ $\mu$m/s \cite{chary1989direct, dafni2002overexpression}, and the speed decreases further with proximity to the cell surface due to the laminar nature of low-Reynolds-number flow. In contrast, a secreted molecule diffuses with coefficient $D = 130$$-$$160$ $\mu$m$^2$/s \cite{fleury2006autologous}, covering a distance equivalent to the cell radius ($a \approx 10$ $\mu$m \cite{shields2007autologous}) in a typical time of $a^2/D$ and giving a ``velocity'' of $D/a = 13$$-$$16$ $\mu$m/s. The ratio of these velocities $\epsilon \equiv v_0a/D = 0.006$$-$$0.08$, called the P\'eclet number, is small, indicating that diffusion dominates over flow in this process.

Also, the secretion rate is low. Cells secrete $0.7$$-$$2.3\times10^{-15}$ g of CCL19/21 ligand in a 24-hour period (Fig.\ 3F in \cite{shields2007autologous}), which given the molecular weights of these ligands ($11$ and $14.6$ kDa, respectively \cite{hornbeck2014phosphositeplus}), corresponds to a secretion rate of $\nu = 1200$$-$$5200$ molecules per hour. Yet, cells begin migrating in a matter of hours \cite{shields2007autologous}.

The slow flow and low secretion rate raise the question of whether autologous chemotaxis is a physically plausible mechanism for these cells. Is a couple thousand molecules, biased by such a weak flow field, enough to determine the flow direction? If so, with what precision? Although this mechanism has been modeled at the continuum level, the question of sensory precision has remained unexplored.

At the same time, the question of sensory precision has been heavily explored for other cellular processes, beginning with the early work of Berg and Purcell \cite{berg1977physics}, and extending to more modern works on concentration sensing \cite{bialek2005physical, wang2007quantifying, endres2009maximum, berezhkovskii2013effect, kaizu2014berg, lang2014thermodynamics, bicknell2015limits, fancher2017fundamental}, gradient sensing \cite{endres2008accuracy, endres2009accuracy, hu2010physical, mugler2016limits, varennes2017emergent}, and related sensory tasks \cite{mora2010limits, siggia2013decisions, mora2015physical, mora2019physical}. Yet, the mechanism of autologous chemotaxis has thus far evaded this list, despite its importance to cancer biology and its potential for interesting physics.

\begin{figure}[b]
\begin{center}
\includegraphics[width=0.48\textwidth]{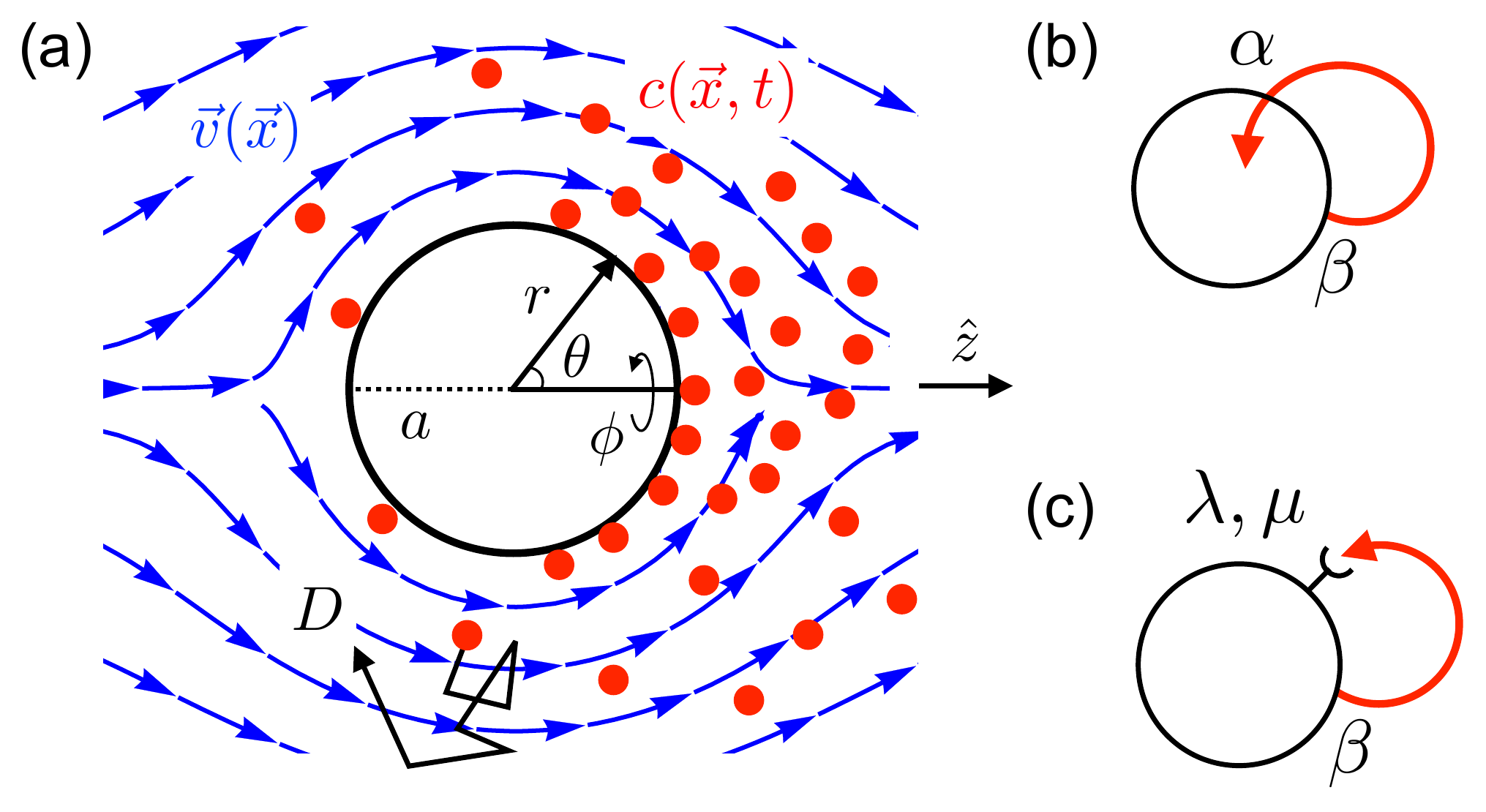}
\end{center}
\caption{Flow sensing by self-communication. (a) A cell isotropically secretes molecules (red) that diffuse and drift along laminar flow lines (blue). The cell detects the molecules by (b) absorption or (c) reversible binding to receptors.}
\label{cartoon}
\end{figure}

Here we combine stochastic techniques from sensory biophysics with perturbation techniques from fluid dynamics to derive the fundamental limit to the precision of flow sensing by self-communication. We consider two ubiquitous methods of molecule detection: absorption vs.\ reversible binding to receptors (Fig.\ \ref{cartoon}). For both, we find a Berg-Purcell-like expression that is ultimately limited by the P{\'e}clet number, the secretion rate, and the integration time. Comparing to the experiments, this expression places a stringent limit on the level of precision that is possible for these cells, suggesting that they detect the flow direction near-optimally given the physical constraints. Finally, we predict that reversible binding is more precise than absorption due to the fact that absorption necessarily reduces the anisotropy in the detected signal, a prediction that we test with endocytosis data on the CCL19/21-CCR7 ligand-receptor pair.

Consider a spherical cell with radius $a$ that secretes molecules isotropically with rate $\beta \equiv \nu/4\pi a^2$ per unit area, in the presence of a fluid flowing with velocity $v_0$ (Fig.\ \ref{cartoon}). At low Reynolds number and high environmental permeability, laminar flow lines obeying Stokes' equation \cite{berg1977physics} form around the cell [Fig.\ \ref{cartoon}(a), blue]. However, in the tumor environment and in experiments, the permeability ${\cal K}$ is low ($\kappa \equiv \sqrt{\cal K}/a \sim 10^{-3}$ \cite{shields2007autologous}), and the flow lines obey the more general Brinkman's equation \cite{brinkman1949calculation}. For a sphere at steady state they are given by \cite{barman1996flow}
\begin{align}
\label{flow}
\vec{v}&(r,\theta,\phi) = v_0\cos\theta
	\left[1- \frac{\zeta}{\rho^3}
  	+ \frac{3\kappa}{\rho^2}\left(1 + \frac{\kappa}{\rho}\right) e^{-(\rho-1)/\kappa} \right] \hat{r} \nonumber\\
&- v_0\sin\theta
	\left[1 +  \frac{\zeta}{2\rho^3}
	- \frac{3}{2\rho}\left(1 + \frac{\kappa}{\rho} + \frac{\kappa^2}{\rho^2}\right) e^{-(\rho-1)/\kappa} \right] \hat{\theta}.
\end{align}
Here, $\rho \equiv r/a$ and $\zeta \equiv 1+3\kappa +3 \kappa^2$, the flow is in the $\hat{z}$ direction ($\theta = 0$), $\hat{r}$ and $\hat{\theta}$ are the radial and polar unit vectors, and $\vec{v}$ is independent of $\phi$ by symmetry. In the limit $\kappa \to \infty$, Eq.\ \ref{flow} reduces to Stokes flow; we are interested in the opposite limit. Note that $\vec{v} = 0$ at the cell surface $r=a$.

The molecules diffuse with coefficient $D$ and drift along the flow lines [Fig.\ \ref{cartoon}(a), red]. This process creates a stochastically evolving concentration field $c(r,\theta,\phi,t)$ with a mean distribution $\bar{c}(r,\theta,\phi,t)$, where the bar represents the ensemble average over many independent realizations of the system. The mean follows the diffusion-drift equation, which at steady state reads
\begin{equation}
\label{dd}
0 = \frac{\partial \bar{c}}{\partial t} = D\nabla^2\bar{c} - \vec{v}\cdot\vec{\nabla}\bar{c}.
\end{equation}
We consider two cases for molecule detection at the cell surface: absorption [Fig.\ \ref{cartoon}(b)] or reversible receptor binding [Fig.\ \ref{cartoon}(c)]. In the former, there exists a flux boundary condition at the cell surface,
\begin{equation}
\label{bc}
-D\left.\frac{\partial \bar{c}(r,\theta)}{\partial r}\right|_a = \beta - \alpha \bar{c}(a,\theta),
\end{equation}
where $\alpha$ is the absorption rate per unit area, and $\bar{c}(r,\theta)$ is independent of $\phi$ and $t$ by symmetry and the system being in steady state, respectively. We also require that the concentration vanish at infinity.

We define the dimensionless concentration $\chi \equiv \bar{c}a^3$ and velocity $\vec{u} \equiv \vec{v}/v_0$. In terms of the dimensionless radial distance $\rho$ and the P\'eclet number $\epsilon$, Eq.\ \ref{dd} at steady state becomes $0 = \nabla_\rho^2\chi - \epsilon\vec{u}\cdot\vec{\nabla}_\rho\chi$. Because $\epsilon$ is small, we use a perturbative solution $\chi = \chi_0 + \epsilon\chi_1$. However, in problems with diffusion and background flow, a single perturbative expansion cannot simultaneously satisfy the boundary conditions at $r=a$ (Eq.\ \ref{bc}) and $r\rightarrow\infty$ ($\bar{c}\to0$) due to the particular spatial nonuniformity of $\vec{u}$ \cite{acrivos1962heat}. The resolution is to split the solution into an inner part $\chi(\rho,\theta)$ that satisfies the boundary condition at the cell surface and holds when $\rho$ is order one, and an outer part $X(s,\theta)$ that satisfies the boundary condition at infinity and holds when $s=\epsilon\rho$ is order one. We match $\chi$ and $X$ by requiring them to be equal at each order in $\epsilon$ as $\rho\rightarrow \infty$ and $s\rightarrow 0$, respectively.

To zeroth order, the inner solution satisfies Laplace's equation, $0 = \nabla_\rho^2\chi_0$, the general solution to which consists of spherical harmonics and powers of $\rho$ \cite{supp}. For the outer solution, we write Eq.\ \ref{dd} in terms of $s$ and $X$, which reads $0 = \nabla_s^2X - \vec{u}\cdot\vec{\nabla}_sX$. One can define a perturbative expansion for $X$, but we show \cite{supp} that only the leading terms of $X$ and $\vec{u}$ matter. The latter is $\vec{u} = \hat{z}$, corresponding to the uniform flow far from the cell where $X$ applies. The solution to this equation satisfying $X\to0$ as $s\to\infty$ consists of modified Bessel functions and spherical harmonics \cite{supp}.

We find that the matching condition requires all but one term in $\chi_0$ and $X$ to vanish \cite{supp}, yielding
\begin{equation}
\label{0}
\chi_0 = \frac{\gamma}{\rho}, \qquad
X = \frac{\epsilon\gamma}{s}e^{-s(1-\cos\theta)/2},
\end{equation}
where $\gamma \equiv \tilde{\beta}/(1+\tilde{\alpha})$, and $\tilde{\beta} \equiv \beta a^4/D$ and $\tilde{\alpha} \equiv \alpha a/D$ are dimensionless secretion and absorption rates, respectively. We see that to leading order, the concentration falls off with distance, and far from the cell it is largest in the flow direction ($\theta = 0$).

To obtain the anisotropy near the cell, which is essential for the flow sensing problem, we must go to the next order. $\chi_1$ satisfies $0 = \nabla_\rho^2\chi_1 - \vec{u}\cdot\vec{\nabla}_\rho\chi_0$, which is the Poisson equation with $\vec{u}$ (Eq.\ \ref{flow}) and $\chi_0$ (Eq.\ \ref{0}) providing the source term. This equation can be solved using a Green's function, with coefficients determined by Eq.\ \ref{bc} and matching to $X$ in Eq.\ \ref{0} \cite{supp}. The result is
\begin{equation}
\label{chi1}
\chi_1 = \frac{\gamma}{2}\left\{\frac{\tilde{\alpha}}{(1+\tilde{\alpha})\rho}-1
	+\frac{\cos\theta}{4}\left[\frac{(1-\tilde{\alpha})w}{(2+\tilde{\alpha})\rho^2}
	+f(\rho,\kappa)\right]\right\},
\end{equation}
where $w \equiv 1 + \kappa^{-1} - \kappa^{-2}e^{1/\kappa}E_1(\kappa^{-1})$ is a monotonic function that limits to $2$ ($\kappa \ll 1$) and $1$ ($\kappa \gg 1$), $f(\rho,\kappa)$ is an $\alpha$-independent function \cite{supp}, and $E_1(x) \equiv \int_1^\infty dt\ e^{-tx}/t$. We see that $\chi_1$ acquires a $\cos\theta$ anisotropy largest in the flow direction ($\theta = 0$). We have checked by numerical solution of Eq.\ \ref{dd} that for $\epsilon \le 0.1$, Eq.\ \ref{chi1} is accurate to within $0.4\%$ at the cell surface \cite{supp, code}.

Information about the anisotropy, and thus the flow direction, comes from the front-back asymmetry in the absorptive flux of molecules $\alpha c$ at the cell surface over a time $T$, which is captured by weighing each absorption event by its location represented as $\cos\theta$. Normalizing this by the mean number of absorbed molecules, we define the anisotropy measure \cite{endres2008accuracy, varennes2017emergent}
\begin{equation}
\label{Aa}
A \equiv \frac{\int_0^T dt \int a^2 d\Omega\ \alpha c(a,\theta,\phi,t) \cos\theta}{T \int a^2 d\Omega'\ \alpha \bar{c}(a,\theta')},
\end{equation}
where $d\Omega =  d\phi\ d\theta\sin\theta$, and the cosine extracts the asymmetry between the front ($\theta = 0$) and back ($\theta = \pi$).
Using the solution for $\chi$ in Eqs.\ \ref{0} and \ref{chi1} and the fact that $f(1,\kappa) = w$, the mean evaluates to \cite{supp}
\begin{equation}
\label{meana}
\bar{A} = \frac{w\epsilon}{8(2+\tilde{\alpha})}
\end{equation}
to leading order in $\epsilon$.

Eq.\ \ref{meana} gives the mean anisotropy but ignores the counting noise due to diffusive molecule arrival. The equivalent expression to Eq.\ \ref{Aa} that accounts for discrete molecule arrival is \cite{endres2008accuracy}
$A = \bar{N}^{-1}\sum_{i=1}^N \cos\theta_i$, where $\theta_i$ is the arrival angle of the $i$th molecule, and $N = \int_0^T dt \int a^2d\Omega\ \alpha c(a,\theta,\phi,t)$ is the total number of molecules absorbed in time $T$.
The mean of this expression is given by Eq.\ \ref{meana} \cite{supp}. The variance is calculated by recognizing that molecule arrivals are statistically independent and that $N$ is Poissonian \cite{endres2008accuracy} (which we have checked even with flow using particle-based simulations \cite{supp, code}). The result is \cite{supp}
\begin{equation}
\label{vara}
\sigma_A^2 = \frac{1}{\bar{N}} = \frac{1}{\nu T}\left(\frac{1+\tilde{\alpha}}{\tilde{\alpha}}\right)
\end{equation}
to leading order in $\epsilon$.
This expression includes (as does Eq.\ \ref{varb} below) a factor of $3$ that arises from each directionally independent component of the variance. We see that the variance in the anisotropy scales inversely with the mean number of absorbed molecules.

Combining Eqs.\ \ref{meana} and \ref{vara}, we obtain a relative error of
\begin{equation}
\label{Ea}
\frac{\sigma^2_A}{\bar{A}^2} = \frac{64(1+\tilde{\alpha})(2+\tilde{\alpha})^2}{w^2\epsilon^2\nu T\tilde{\alpha}}
	\gtrsim \frac{282}{\epsilon^2\nu T}.
\end{equation}
In the second step, we have set $w$ to its maximal value of $2$ for $\kappa \ll 1$ (as in the experiments \cite{shields2007autologous}) and recognized that the expression has a minimum at $\tilde{\alpha}^* = (\sqrt{17}-1)/4 \approx 0.78$. The minimum arises from the following tradeoff: strong absorption maximizes the number of detected molecules and therefore reduces noise (Eq.\ \ref{vara}); but it also causes molecules to be absorbed immediately after release, preventing them from interacting with the nonzero flow away from the cell surface and therefore reducing the mean (Eq.\ \ref{meana}). Eq.\ \ref{Ea} sets the fundamental limit to the precision of flow sensing by molecule absorption, dependent only on the P\'eclet number $\epsilon$ and the total number of secreted molecules $\nu T$.

We now consider the case of reversible receptor binding [Fig.\ \ref{cartoon}(c)]. Calling $b(\theta, \phi, t)$ the surface concentration of bound receptors, we have
\begin{align}
\frac{\partial c}{\partial t} &= D\nabla^2c - \vec{v}\cdot\vec{\nabla}c + \eta_D
	+ \left(-\frac{\partial b}{\partial t} + \beta + \eta_\beta \right)\delta(r-a), \nonumber \\
\label{cb}
\frac{\partial b}{\partial t} &= \lambda c(a,\theta,\phi,t) - \mu b + \eta_b,
\end{align}
where the term proportional to the delta function contains the boundary condition at the surface.
Here $\lambda \equiv k_a(R/4\pi a^2 - b) \approx k_aR/4\pi a^2$ and $\mu$ are the binding and unbinding rates, respectively, where $k_a$ is the intrinsic ligand-receptor association rate, and $R$ is the number of receptors per cell.
Because binding is reversible, there are correlations between the bound receptor concentrations at different regions of the cell surface. Therefore, we cannot use the Poisson counting technique (Eq.\ \ref{vara}) to calculate the noise. Instead, we include Langevin noise terms in Eq.\ \ref{cb} to account for these correlations. These terms have zero mean, are uncorrelated with each other, and satisfy \cite{gardiner2004handbook, gillespie2000chemical, fancher2017fundamental, varennes2017emergent}
\begin{align}
\langle \eta_D(\vec{r},t)\eta_D(\vec{r}\ ',t')\rangle
	&= 2D\delta(t-t')\vec{\nabla}_r\cdot\vec{\nabla}_{r'}[\bar{c}(\vec{r})\delta(\vec{r}-\vec{r}\ ')], \nonumber \\
\langle \eta_\beta(\Omega,t) \eta_\beta(\Omega',t')\rangle &= \beta \delta(\Omega-\Omega')\delta(t-t'), \\
\langle \eta_b(\Omega,t) \eta_b(\Omega',t')\rangle &= 2\mu\bar{b}\ \delta(\Omega-\Omega')\delta(t-t'), \nonumber
\end{align}
where $\bar{c}(r,\theta)$ and $\bar{b}(\theta) = \lambda\bar{c}(a,\theta)/\mu$ are the mean concentrations in steady state. Binding and unbinding equilibrate in steady state, such that $\bar{c}(r,\theta)$ is given by the previous solution (Eqs.\ \ref{0} and \ref{chi1}) but with $\alpha = 0$. The approximation in the definition of $\lambda$ above neglects receptor saturation, which is valid because $\bar{c}(a)/K_d = \nu/4\pi aDK_d \sim 10^{-4}$, where we have used the isotropic approximation for $\bar{c}(a)$ (Eq.\ \ref{0}, $\alpha = 0$) and a dissociation constant of $K_d = \mu/k_a \sim 1$ nM for the CCL19/21-CCR7 ligand-receptor pair \cite{willimann1998chemokine, yoshida1998secondary}.

In the reversible binding case, the anisotropy is defined as the average of the cosine over the angular distribution of bound receptors and the integration time $T$,
\begin{equation}
\label{Ab}
A \equiv \frac{\int_0^T dt \int a^2d\Omega\ b(\theta,\phi,t)\cos\theta}{T \int a^2 d\Omega'\ \bar{b}(\theta')}.
\end{equation}
Because $\bar{b}(\theta) = \lambda\bar{c}(a,\theta)/\mu$, the means of Eqs.\ \ref{Aa} and \ref{Ab} take equivalent forms. Therefore, to leading order in $\epsilon$, the mean of Eq.\ \ref{Ab} is simply Eq.\ \ref{meana} with $\alpha = 0$,
\begin{equation}
\label{meanb}
\bar{A} = \frac{w\epsilon}{16}.
\end{equation}
To solve Eqs.\ \ref{cb}-\ref{Ab} for the variance, we Fourier transform them in space and time, calculate the power spectrum of $A$, and recognize that $\sigma_A^2T$ is given by its low-frequency limit \cite{bialek2005physical, mugler2016limits, fancher2017fundamental, varennes2017emergent}. The result is \cite{supp}
\begin{equation}
\label{varb}
\sigma_A^2 = \frac{1}{\nu T}\left(\frac{7}{9} + \frac{2}{\tilde{\lambda}}\right)
\end{equation}
to leading order in $\epsilon$, where $\tilde{\lambda} \equiv \lambda a/D$. The two terms are from noise due to (i) secretion and diffusion, and (ii) binding and unbinding, respectively. The derivation of Eq.\ \ref{varb} assumes that $T \gg \{\tau_1, \tau_2\}$, where $\tau_1\equiv a^2/D \sim 1$ s is the characteristic time for a ligand molecule to diffuse across the cell, and $\tau_2 \equiv (1+\tilde{\lambda})/\mu \approx \tilde{\lambda}/\mu = R/4\pi aDK_d \sim 1$$-$$10$ s is the receptor equilibration timescale \cite{fancher2017fundamental}. For $\tau_2$ we take $R \sim 10^4$$-$$10^5$ CCR7 receptors per cell \cite{willimann1998chemokine, comerford2006chemokine} and $\tilde{\lambda} \gg 1$, which corresponds to diffusion-limited binding as further discussed below. Because cells migrate over hours, we see that $T \gg \{\tau_1, \tau_2\}$ should indeed be valid.

Combining Eqs.\ \ref{meanb} and \ref{varb}, we obtain the relative error
\begin{equation}
\label{Eb}
\frac{\sigma_A^2}{\bar{A}^2} = \frac{1792}{9w^2\epsilon^2\nu T} \left(1 + \frac{18}{7\tilde{\lambda}}\right)
	\gtrsim \frac{50}{\epsilon^2\nu T}.
\end{equation}
In the second step, we again take $w = 2$ and $\tilde{\lambda} \gg 1$. Comparing Eqs.\ \ref{Ea} and \ref{Eb}, we see that reversible binding achieves $\sqrt{282/50} \approx 2.4$ times lower error than absorption. The reason is that absorption (Eq.\ \ref{meana}), but not binding (Eq.\ \ref{meanb}), reduces the anisotropy. Absorption is an active modifier of the signal created by secretion and flow, whereas reversible binding is a passive monitor.

\begin{figure}
\begin{center}
\includegraphics[width=0.4\textwidth]{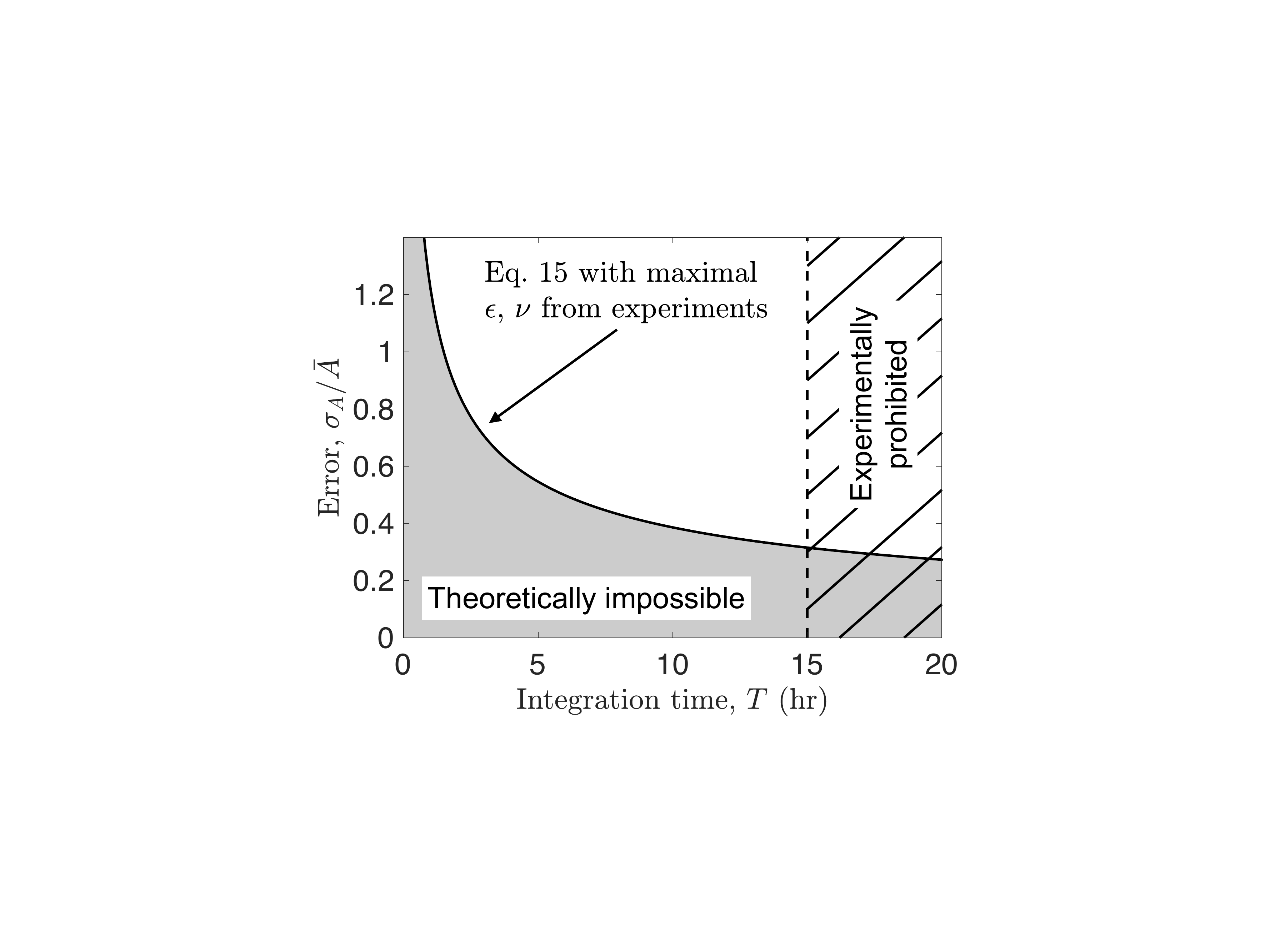}
\end{center}
\caption{Fundamental limit to the precision of flow sensing. Maximum experimental values $\epsilon = 0.08$ and $\nu = 5200$/hr \cite{shields2007autologous} are used for minimum error (solid line). Cells migrate within 15 hours  \cite{shields2007autologous} (dashed line). Lowest possible error is 30\%.}
\label{limit}
\end{figure}

How do our results compare to the experiments on metastatic cancer cells? The inequality in Eq.\ \ref{Eb} provides the fundamental detection limit. We plot this expression as a function of $T$ in Fig.\ \ref{limit} using the maximal experimental values of $\epsilon = 0.08$ and $\nu = 5200$/hr \cite{shields2007autologous} to obtain the minimum possible error. We see that low errors are not possible in a few hours; even 10\% error would take over 150 hours to achieve. Yet, the cells are observed to migrate over a 15 hour period \cite{shields2007autologous}. In this time frame, it is not possible to achieve less than 30\% error (Fig.\ \ref{limit}). The situation is likely worse, given that the cells presumably begin migrating well before the 15-hour mark, and given that we have neglected any internal signaling noise. Thus, we see that the sensory performance is severely limited by the experimental parameters and the physics of the detection process. We conclude that these cells operate remarkably close to the fundamental detection limit.

We find that absorption is less precise than reversible binding (Eqs.\ \ref{Ea} and \ref{Eb}). A ubiquitous mechanism of ligand absorption is endocytosis, wherein bound receptors are internalized into the cell. Therefore, we predict that the degree of CCR7 endocytosis in response to CCL19/21 binding is low. This prediction can be tested with endocytosis data on this ligand-receptor pair. Specifically, to achieve optimal absorption in Eq.\ \ref{Ea} ($\tilde{\alpha}^* \approx 0.78$), absorption would need to occur at a rate of $4\pi a^2 \alpha^* \bar{c}(a) = \nu\tilde{\alpha}^*/(1+\tilde{\alpha}^*) \sim 25$ min$^{-1}$, where we have used the isotropic approximation for $\bar{c}(a)$ (Eq.\ \ref{0}). However, the rate of CCR7 endocytosis in response to CCL19/21 binding is many times slower at about $1$ min$^{-1}$ \cite{byers2008arrestin}. Thus, the degree of endocytosis is much lower than required for the absorption mechanism, as predicted.

We also find that reversible binding is most precise when the parameter $\tilde{\lambda} = Rk_a/4\pi aD$ is large (Eq.\ \ref{Eb}). Writing this parameter as $\tilde{\lambda} = (k_a/4\pi\ell D)(R\ell/a)$, where $\ell$ is the receptor lengthscale, we see that the first factor is the ratio that determines whether ligand-receptor binding is diffusion-limited ($k_a \gg 4\pi\ell D$) or reaction-limited ($k_a \ll 4\pi\ell D$). With the known values of $R$ and $a$ and a typical receptor lengthscale of $\ell \sim 10$ nm, the second factor evaluates to $10$$-$$100$. Therefore, the requirement that $\tilde{\lambda} \gg 1$ is equivalent to the statement that binding is either diffusion-limited or weakly reaction-limited. Given the high sensory performance implied by Fig.\ \ref{limit} and the low degree of endocytosis found above, we thus predict that CCL19/21 binding to CCR7 is either diffusion-limited or weakly reaction-limited. We are not aware of kinetics data that would test this prediction.

Our finding that reversible binding is more precise than absorption is the opposite of what was found for the detection of an externally established concentration gradient \cite{endres2008accuracy}. The reason is that in our problem absorption removes molecules at the source, whereas in that problem molecules are replenished by a source at infinity. Depletion at the source prevents interactions with the flow and therefore weakens the anisotropy. Additionally, our models do not include any additional noise sources from processes internal to the cell such as protein signaling or gene expression. Because any such process would simply add a fixed amount of noise, our finding is unaffected by the inclusion of internal dynamics, and Eq.\ \ref{Eb} remains a theoretical minimum to the error in flow sensing.

The severity of the limit in Fig.\ \ref{limit} raises the question of whether metastatic cancer cells benefit from additional sensory mechanisms not accounted for in our modeling. The precision of flow sensing may be affected by geometric properties of the cell such as a nonuniform distribution of receptors or aspherical morphology. We find that receptor clustering has a negligible effect on the anisotropy but that an ellipsoidal cell \cite{obrien1965eggs, frankel2011geometry} can decrease its sensory error by elongating in the direction of the flow \cite{supp, code}. Further investigation of the effects of cell geometry would be an interesting topic for future work. Some chemoattractants including CCL21 are known to bind to extracellular matrix fibers and be subsequently released by proteases \cite{patel2001chemokines, sahni1998binding, sahni2004interleukin, lee2005processing}. This effect has been shown in continuum models of autologous chemotaxis to substantially increase the anisotropy \cite{helm2005synergy, fleury2006autologous}, although the impact on the noise is unknown. It is also important to recognize that these cells do not perform flow sensing in isolation. Indeed, studies have shown that their migration is (i) increased in the presence of another cell type (fibroblasts) \cite{shieh2011tumor}, (ii) decreased at high cell densities \cite{polacheck2011interstitial}, and (iii) reversed at even higher cell densities (although reversal is attributed to a separate pressure-sensing mechanism) \cite{polacheck2011interstitial}. The extension of our work to multiple cells remains to be explored. Finally, recent work has highlighted the benefit of on-the-fly sensing \cite{siggia2013decisions, desponds2019hunchback}, where an agent makes (and continually updates) its decision during the integration time, instead of afterward as assumed here. On-the-fly sensing may play an important role for these cells.

We have derived the fundamental limit to flow sensing by self-communication and shown that it strongly constrains the performance of metastatic cancer cells. Our work elucidates the physics behind a fascinating detection process and provides quantitative insights into a critical step in cancer progression.

\acknowledgments
This work was supported by Simons Foundation grant 376198 and National Science Foundation grant MCB-1936761.
We thank Nicholas Licata for useful discussions.


\onecolumngrid
\section{Supplemental Material}

\section{Derivation of Eq.\ 4}\label{ZeroOrderSolution}

In this section, we derive the lowest order terms in the expansion for the inner and outer solution (Eq.\ 4 of the main text). We recall the non-dimensionalized variables and parameters
\begin{equation}\label{eq:nondimabs}
    \begin{gathered}
        \chi = \bar{c} a^3, \qquad \rho = \frac{r}{a}, \qquad \tilde{\beta} = \frac{\beta a^4}{D}, \\
        \tilde{\alpha} = \frac{\alpha a}{D}, \qquad \epsilon = \frac{v_0 a}{D}, \qquad \kappa = \frac{\sqrt{\mathcal{K}}}{a}.
    \end{gathered}
\end{equation}
It will be convenient to describe the flow profile with the functions 
\begin{equation} \label{ureqB}
    \begin{gathered}
        u_r(\rho) = 1 -\frac{1 + 3\kappa +3 \kappa^2}{\rho^3} + \frac{3\kappa}{\rho^2}\left( 1+ \frac{\kappa}{\rho} \right) e^{\frac{1-\rho}{\kappa}}, \\
        u_{\theta}(\rho) = 1 + \frac{1 + 3\kappa +3 \kappa^2}{2\rho^3} -\frac{3}{2\rho}\left(1+\frac{\kappa}{\rho}+\frac{\kappa^2}{\rho^2} \right) e^{\frac{1-\rho}{\kappa}}
    \end{gathered}
\end{equation}
With these, the dimensionless flow profile (Eq.\ 1 of the main text) is
\begin{equation}\label{eq:BrinkmannFlow}
    \vec{u}(\rho,\theta) = \frac{\vec{v}(\rho,\theta)}{v_0} = u_r(\rho) \cos\theta \hat{\rho} - u_{\theta}(\rho) \sin\theta \hat{\theta}.
\end{equation}

We solve the drift-diffusion equation (Eq.\ 2 of the main text) with these flow lines through the method of matched asymptotic expansions. To do this, we introduce two expansions: an inner and an outer one. The inner one satisfies the boundary condition at the cell surface, while the outer one satisfies the condition at infinity. We obtain the full solution and remaining coefficients by matching the functional forms on a common overlap region: $s=\epsilon \rho \rightarrow 0$ for the outer expansion and $\rho\rightarrow \infty$ for the inner expansion.

We assume that the inner expansion has the standard form
\begin{equation}
    \chi(\rho,\theta,\epsilon) = \sum\limits_{n=0}^{\infty} \epsilon^n \chi_n(\rho,\theta).
\end{equation}
This is a solution to the problem
\begin{equation}
    0 = \nabla_{\rho}^2 \chi(\rho,\theta) - \epsilon \vec{u}(\rho,\theta)\cdot \nabla_{\rho}\chi(\rho,\theta), \quad -\frac{\partial \chi(\rho,\theta)}{\partial \rho} \Big|_{\rho=1} = \tilde{\beta} - \tilde{\alpha} \chi(1,\theta),
\end{equation}
where $\chi$'s $\epsilon$ dependence has been suppressed. Collecting powers of $\epsilon$, the equations for $\chi_n$ become
\begin{equation}\label{eq:innerProb}
    0 = \nabla_{\rho}^2 \chi_{n} - \vec{u} \cdot\nabla_{\rho} \chi_{n-1}, \qquad -\frac{\partial \chi_n}{\partial \rho}\Big|_{\rho=1} = \tilde{\beta} \delta_{n,0} - \tilde{\alpha} \chi_n(1).
\end{equation}
This assumes that the flow is small, which is valid close to the surface of the cell.

For the outer expansion, we introduce the re-scaled distance
\begin{equation}
    s= \epsilon \rho.
\end{equation}
We make the standard choice
\begin{equation}
    X(s,\theta,\epsilon) = \sum\limits_{n=0}^{\infty} F_n(\epsilon) X_n(s,\theta),
\end{equation}
where
\begin{equation}
    \lim\limits_{\epsilon\rightarrow 0} \frac{F_{n+1}(\epsilon)}{F_n(\epsilon)}=0.
\end{equation}
In the derivation of Eq.\ 5 of the main text (next section), we will show that using only the lowest order term $F_0(\epsilon)X_0(s,\theta)$ gives a consistent solution sufficient for our purposes. The full expansion solves the problem
\begin{equation}\label{eq:fullX}
    0 = \nabla_s^2 X(s,\theta) - \vec{u}\left( \frac{s}{\epsilon},\theta \right) \cdot \nabla_s X(s,\theta), \quad \lim\limits_{s\rightarrow \infty} X_n(s,\theta) = 0.
\end{equation}
For the outer expansion, we neglect the exponential terms in $\vec{u}$, as these have $-s/\epsilon$ in the exponent, which is smaller than any power of $\epsilon$ and cannot be captured by a Taylor series. This means that we work with
\begin{equation}
        u_r\left(\frac{s}{\epsilon} \right) \sim 1 -\frac{\zeta \epsilon^3}{s^3} ,\qquad u_{\theta}\left(\frac{s}{\epsilon} \right) \sim 1 + \frac{2\zeta \epsilon^3}{s^3},
\end{equation}
where $\zeta = 1 + 3\kappa +3 \kappa^2$. For the lowest order terms (order $0$$-$$2$), only the constant terms in the $\vec{u}$ affect the PDE, and the flow is just the flow at infinity, $\hat{z}$.

\subsection{Inner Expansion}

For the zero-order term, we have
\begin{equation}\label{eq:innerProb2}
    \nabla_{\rho}^2 \chi_0 = 0, \qquad -\left.\frac{\partial \chi_0}{\partial \rho}  \right|_{\rho=1} = \tilde{\beta} -\tilde{\alpha} \chi_0(1).
\end{equation}
Using azimuthal symmetry, the general solution to the PDE is
\begin{equation}\label{eq:chi0gen}
    \chi_0 = \sum\limits_{\ell = 0}^{\infty} \left( A_{0,\ell} \rho^{\ell} + \frac{B_{0,\ell}}{\rho^{\ell+1}} \right) Y_{\ell}^0(\theta),
\end{equation}
where $A_{0,\ell}$ and $B_{0,\ell}$ are undetermined coefficients, and $Y_\ell^m$ are spherical harmonics.
We plug this into the boundary condition in Eq.\ \ref{eq:innerProb2}. Since the spherical harmonics are linearly independent, we have the system
\begin{equation}\label{eq:zeroOrderInnerBC}
    -(\ell A_{0,\ell} -(\ell+1)B_{0,\ell}) = \sqrt{4\pi} \tilde{\beta}\delta_{0,\ell} - \tilde{\alpha}(A_{0,\ell} +B_{0,\ell}),
\end{equation}
where the factor of $\sqrt{4\pi}$ arises from $Y_0^0=(4\pi)^{-1/2}$ and factoring off a spherical harmonic from both sides of the equation.
We will use this result shortly, as it will simplify substantially after using the matching condition.

\subsection{Outer Expansion}

The equation for $X_0$ follows from Eq.\ \ref{eq:fullX},
\begin{equation}\label{eq:zeroOuterPDE}
    0=\nabla_s^2 X_0 -\cos\theta \frac{\partial X_0}{\partial s} +\frac{\sin\theta}{s} \frac{\partial X_0}{\partial \theta},
\end{equation}
where as discussed we use $\vec{u} = \hat{z}$ and we have written the gradient in spherical coordinates.
We can eliminate the $\theta$-dependence and replace it with a $\cos\theta$-dependence using
\begin{equation}
    \sin\theta \frac{\partial }{\partial \theta} = -(1-\cos^2\theta)  \frac{\partial }{\partial (\cos\theta)},
\end{equation}
with which  Eq.\ \ref{eq:zeroOuterPDE} becomes
\begin{equation}
    0=\nabla_s^2 X_0 -\cos\theta \frac{\partial X_0}{\partial s} -\frac{(1-\cos^2\theta)}{s} \frac{\partial X_0}{\partial (\cos\theta)}. 
\end{equation}
If we make the subsitution $X_0(s,\theta) =G(s,\theta)\exp(s\cos(\theta)/2)$, the equation simplifies because the operator becomes
\begin{equation}
    \nabla_s^2 X_0 -\cos\theta \frac{\partial X_0}{\partial s} -\frac{(1-\cos^2\theta)}{s} \frac{\partial X_0}{\partial (\cos\theta)} = e^{\frac{s}{2}\cos\theta} \left[\nabla_s^2 -\frac{1}{4} \right]G(s,\theta).
\end{equation}
Since the exponential factor never vanishes, the PDE becomes
\begin{equation}
    \nabla_s^2 G(s,\theta) - \frac{1}{4}G(s,\theta)=0.
\end{equation}
To move forward, we write $G$ as a linear combination of spherical harmonics and use azimuthal symmetry
\begin{equation}
    G(s,\theta) =\sum\limits_{\ell= 0}^{\infty} \frac{H_{\ell}(s/2)}{\sqrt{s}} Y_{\ell}^0(\theta),
\end{equation}
where the $H_\ell$ are to be determined. Substitution and isolating the independent spherical harmonics give the ODEs
\begin{equation}
    0 = s^{-5/2} \left[ \left( \frac{s}{2} \right)^2 \frac{d^2 H_{\ell}(\frac{s}{2})}{d(\frac{s}{2})^2} + \frac{s}{2}  \frac{d H_{\ell}(\frac{s}{2})}{d(\frac{s}{2})} -\left(\left( \frac{s}{2} \right)^2 + \left(\ell + \frac{1}{2} \right)^2 \right)H_{\ell}\left(\frac{s}{2}\right) \right].
\end{equation}
The term in square brackets must vanish, and this is just the modified Bessel differential equation in $s/2$ of order $\ell+1/2$. This means that the general solution for $H_{\ell}$ is
\begin{equation}
    H_{\ell}(s/2) = C_{0,\ell}K_{\ell+1/2}(s/2) + D_{0,\ell}I_{\ell+1/2}(s/2),
\end{equation}
where the $I$s and $K$s are modified Bessel functions of the first and second kind, respectively, and $C_{0,\ell}$ and $D_{0,\ell}$ are undetermined coefficients. Substituting this back into $X$ gives
\begin{equation}
    X_0(s,\theta) = \frac{e^{\frac{s}{2}\cos\theta}}{\sqrt{s}} \sum\limits_{\ell= 0}^{\infty} \left[C_{0,\ell}K_{\ell+1/2}(s/2) + D_{0,\ell}I_{\ell+1/2}(s/2) \right]Y_{\ell}^{0}(\theta).
\end{equation}
Since $X_0$ must vanish at infinity, we must have $D_{0,\ell} = 0$ for all $\ell$ so
\begin{equation}
    X_0(s,\theta) = \frac{e^{\frac{s}{2}\cos\theta}}{\sqrt{s}} \sum\limits_{\ell= 0}^{\infty} C_{0,\ell}K_{\ell+1/2}(s/2)Y_{\ell}^{0}(\theta).
\end{equation}
For positive half-integer orders, the Bessel $K$s are exponentially decaying functions with decaying power laws. The exponentially decaying factor is $\exp(-s/2)$, so the combination of the two exponentials is decreasing for all $\theta$ values except $\theta=0$, where the factor is constant.

\subsection{Asymptotic Matching}

Now we match the functional forms of the two solutions. We look at the inner expansion first (Eq.\ \ref{eq:chi0gen}). Each term in the outer expansion decreases as $s$ increases, so we cannot have the positive powers of $\rho$ in the inner expansion. This implies that $A_{0,\ell}=0$ for $\ell\geq 1$. Since $\tilde{\alpha}\geq 0$, applying the surface boundary condition in Eq.\ \ref{eq:zeroOrderInnerBC} also gives $B_{0,\ell}=0$ for $\ell\geq 1$. This means that, to lowest order in $\epsilon$
\begin{equation}\label{eq:fullChiZeroOrder}
    \chi_0 = Y_{0}^{0}\left( A_{0,0} + \frac{B_{0,0}}{\rho}\right).
\end{equation}

Now we turn to the outer expansion. Note that the modified Bessel functions of the second kind $K$ have the following asymptotics
\begin{equation}
    K_{\ell+1/2}(s/2) = \mathcal{O}(s^{-(\ell+1/2)}), \qquad s\rightarrow 0.
\end{equation}
Including the overall factor of $s^{-1/2}$, we see that the $\ell$ term diverges like $s^{-\ell-1}$. This means that all terms with $\ell>0$ diverge faster than the inner solution, so the coefficients for these terms must be zero, because they cannot be matched. This means
\begin{equation}
    X_0(s,\theta) = \frac{e^{\frac{s}{2}\cos\theta}}{\sqrt{s}}  C_{0,0}K_{1/2}(s/2)Y_{0}^{0}= \sqrt{\pi} C_{0,0} \frac{e^{\frac{s}{2}(\cos\theta-1)}}{s}Y_{0}^{0}.
\end{equation}
and therefore to lowest order in $\epsilon$ we have
\begin{equation}\label{eq:fullXZeroOrder}
    X =  F_0(\epsilon) \sqrt{\pi}C_{0,0} \frac{e^{\frac{s}{2}(\cos\theta-1)}}{s} Y_0^0.
\end{equation}

So far, we have used matching to argue which terms should vanish. Now we will find the values for the non-zero coefficients. To do this, we recognize that because $s = \epsilon \rho$ in Eq.\ \ref{eq:fullXZeroOrder},
in order to match this with Eq.\ \ref{eq:fullChiZeroOrder} in powers of $\epsilon$, we must take
\begin{equation}
    F_0(\epsilon) = \epsilon, \qquad A_{0,0} = 0.
\end{equation}
Using the boundary condition at the surface from Eq.\ \ref{eq:zeroOrderInnerBC} gives
\begin{equation}\label{eq:b00}
    B_{0,0} = \sqrt{4\pi} \frac{\tilde{\beta}}{1+\tilde{\alpha}} = \sqrt{4\pi} \gamma,
\end{equation}
where we define $\gamma = \tilde{\beta}/(1+\tilde{\alpha})$.
Matching Eq.\ \ref{eq:fullXZeroOrder} to the $\rho^{-1}$ term in Eq.\ \ref{eq:fullChiZeroOrder} then gives
\begin{equation}
    C_{0,0} = 2\gamma.
\end{equation}
Using the values determined in this section, Eqs.\ \ref{eq:fullChiZeroOrder} and \ref{eq:fullXZeroOrder} become
\begin{equation}
\label{0supp}
\chi_0 = \frac{\gamma}{\rho}, \qquad
X = \frac{\epsilon\gamma}{s}e^{-s(1-\cos\theta)/2},
\end{equation}
as in Eq.\ 4 of the main text.

\section{Derivation of Eq.\ 5}\label{FirstOrderChiCalc}

In this section, we will calculate the first-order term in the inner expansion and show that we just need the lowest order term in the outer expansion.

\subsection{Inner Expansion}

The first-order term in the inner expansion solves the PDE
\begin{equation}\label{eq:firstOrderInnerPDE}
0 = \nabla_{\rho}^2 \chi_1 - \vec{u} \cdot\nabla_{\rho} \chi_{0}, \qquad -\left.\frac{\partial \chi_1}{\partial \rho}  \right|_{\rho=1} = - \tilde{\alpha} \chi_1(1).
\end{equation}
Using the zero-order solution $\chi_0$ gives
\begin{equation}
    \nabla_{\rho}^2 \chi_1 = \vec{u} \cdot\nabla_{\rho} \chi_{0} = u_r(\rho) \cos(\theta) \left(-\frac{\gamma}{\rho^2} \right) = - \sqrt{\frac{4\pi}{3}}\frac{\gamma}{\rho^2} u_r(\rho)Y_1^0(\theta).
\end{equation}
The general solution to this is the solution to the homogeneous equation (Laplace's equation) plus an inhomogeneous term arising from the presence of a source (the particular solution). We proceed by using the Green's function for Laplace's equation
\begin{equation}\label{eq:green}
    \nabla_{\rho}G(\vec{\rho},\vec{\rho}\text{ }') = \delta^3(\vec{\rho}-\vec{\rho}\text{ }') \Longrightarrow G(\vec{\rho},\vec{\rho}\text{ }') = -\frac{1}{4\pi |\vec{\rho}-\vec{\rho}\text{ }'|}.
\end{equation}
The particular solution is the convolution of this with the source term,
\begin{equation}
\chi_{1}\left(\vec{\rho}\right) = \int_{\rho'\ge 1}d^{3}\rho'G\left(\vec{\rho},\vec{\rho}'\right)\left(-\sqrt{\frac{4\pi}{3}}\frac{\gamma}{{\rho'}^{2}}u_{r}\left(\rho'\right)Y_{1}^{0}\left(\theta'\right)\right).
\label{chi1Gform}
\end{equation}
We expand the Green's function in terms of Legendre polynomials $P_{\ell}$
\begin{equation}
    \frac{1}{|\vec{\rho}-\vec{\rho}\text{ }'|} = \frac{1}{\rho_>} \sum\limits_{\ell=0}^{\infty} \left( \frac{\rho_<}{\rho_>}\right)^{\ell} P_{\ell}(\hat{\rho}\cdot \hat{\rho}'),
\end{equation}
where $\rho_< = \text{min}(\rho,\rho')$ and $\rho_> = \text{max}(\rho,\rho')$. The Legendre polynomials are related to the spherical harmonics via
\begin{equation}\label{eq:legendre2Spherical}
    P_{\ell}(\hat{\rho}\cdot \hat{\rho}') = \frac{4\pi}{2\ell +1} \sum\limits_{m=-\ell}^{\ell} Y_{\ell}^m(\hat{\rho})Y_{\ell}^m(\hat{\rho}')^*.
\end{equation}
By orthogonality, only the term with $\ell=1$ and $m=0$ will make a non-vanishing contribution to the convolution. To evaluate the convolution, we use the orthogonality of spherical harmonics to simplify the angular integrals and break the integral over $\rho'$ into regions where $\rho'<\rho$ and $\rho'>\rho$. Specifically, combining Eqs.\ \ref{eq:green}-\ref{eq:legendre2Spherical} allows the angular portion of the integral to be easily performed,
\begin{align}
\chi_{1}\left(\vec{\rho}\right) &= \int_{\rho'\ge 1}d^{3}\rho'\left(\sum_{\ell,m}\frac{\rho_{<}^{\ell}}{\rho_{>}^{\ell+1}\left(2\ell+1\right)}Y_{\ell}^{m}\left(\hat{\rho}\right)Y_{\ell}^{m*}\left(\hat{\rho}'\right)\right)\left(\sqrt{\frac{4\pi}{3}}\frac{\gamma}{{\rho'}^{2}}u_{r}\left(\rho'\right)Y_{1}^{0}\left(\theta'\right)\right) \nonumber\\
&= \frac{\gamma}{3}\sqrt{\frac{4\pi}{3}}Y_{1}^{0}\left(\theta\right)\left\lbrack\int_{1}^{\rho}d\rho'\frac{\rho'}{\rho^{2}}u_{r}\left(\rho'\right)+\int_{\rho}^{\infty}d\rho'\frac{\rho}{{\rho'}^{2}}u_{r}\left(\rho'\right)\right\rbrack.
\label{chi1Grhoint}
\end{align}
Inserting the expression for $u_r$ (Eq.\ \ref{ureqB}) with $\zeta = 1 + 3\kappa +3 \kappa^2$ into Eq.\ \ref{chi1Grhoint} then yields
\begin{align}
\chi_{1}\left(\vec{\rho}\right) &= \frac{\gamma}{3}\sqrt{\frac{4\pi}{3}}Y_{1}^{0}\left(\theta\right)\left\lbrack\int_{1}^{\rho}d\rho'\frac{\rho'}{\rho^{2}}\left(1-\frac{\zeta}{{\rho'}^{3}}+\frac{3\kappa}{{\rho'}^{2}}\left(1+\frac{\kappa}{\rho'}\right)e^{\frac{1-\rho'}{\kappa}}\right)
+\int_{\rho}^{\infty}d\rho'\frac{\rho}{{\rho'}^{2}}\left(1-\frac{\zeta}{{\rho'}^{3}}+\frac{3\kappa}{{\rho'}^{2}}\left(1+\frac{\kappa}{\rho'}\right)e^{\frac{1-\rho'}{\kappa}}\right)\right\rbrack \nonumber\\
&= \frac{\gamma}{3}\sqrt{\frac{4\pi}{3}}Y_{1}^{0}\left(\theta\right)\left\lbrack\frac{3}{2}-\frac{2\zeta+1}{2\rho^{2}}+\frac{3\zeta}{4\rho^{3}}+\int_{1}^{\infty}d\rho'\frac{3\kappa}{\rho^{2}\rho'}\left(1+\frac{\kappa}{\rho'}\right)e^{\frac{1-\rho'}{\kappa}}
+\int_{\rho}^{\infty}d\rho'\frac{3\kappa}{{\rho'}^{2}}\left(\frac{\rho}{{\rho'}^{2}}-\frac{\rho'}{\rho^{2}}\right)\left(1+\frac{\kappa}{\rho'}\right)e^{\frac{1-\rho'}{\kappa}}\right\rbrack \nonumber\\
&= \frac{\gamma}{2}\sqrt{\frac{4\pi}{3}}Y_{1}^{0}\left(\theta\right)\left\lbrack 1-\frac{2\zeta+1}{3\rho^{2}}+\frac{\zeta}{2\rho^{3}}+\frac{2\kappa}{\rho^{2}}e^{\frac{1}{\kappa}}\left(E_{1}\left(\frac{1}{\kappa}\right)+\kappa E_{2}\left(\frac{1}{\kappa}\right)\right)\right. \nonumber\\
&\quad\quad\quad\quad\quad\quad\quad\quad +\left.\frac{2\kappa}{\rho^{2}}e^{\frac{1}{\kappa}}\left(E_{4}\left(\frac{\rho}{\kappa}\right)-E_{1}\left(\frac{\rho}{\kappa}\right)+\frac{\kappa}{\rho}E_{5}\left(\frac{\rho}{\kappa}\right)-\frac{\kappa}{\rho}E_{2}\left(\frac{\rho}{\kappa}\right)\right)\right\rbrack,
\label{chi1GB}
\end{align}
where
\begin{equation}
E_{n}\left(x\right) = \int_{1}^{\infty}dt\frac{e^{-tx}}{t^{n}}.
\label{Endef}
\end{equation}
Eq. \ref{chi1GB} can be simplified slightly using the recursion relation
\begin{equation}
E_{n}\left(x\right) = \frac{1}{n-1}\left(e^{-x}-xE_{n-1}\left(x\right)\right),
\label{Enrec}
\end{equation}
which is valid for $x>0$. For integer values of $n>1$, this relation can be repeated to produce
\begin{equation}
E_{n}\left(x\right) = \frac{1}{\left(n-1\right)!}\left\lbrack\left(-x\right)^{n-1}E_{1}\left(x\right)+e^{-x}\sum_{i=0}^{n-2}\left(\left(n-2-i\right)!\right)\left(-x\right)^{i}\right\rbrack.
\label{Enreprec}
\end{equation}
Applying these relations to the $E_{n}$ functions seen in Eq. \ref{chi1GB} allows for the simplifications
\begin{align}
E_{1}\left(\frac{1}{\kappa}\right)+\kappa E_{2}\left(\frac{1}{\kappa}\right) &= E_{1}\left(\frac{1}{\kappa}\right)+\kappa\left(e^{-\frac{1}{\kappa}}-\frac{1}{\kappa}E_{1}\left(\frac{1}{\kappa}\right)\right) \nonumber\\
&= \kappa e^{-\frac{1}{\kappa}},
\label{Ensim1}
\end{align}
and
\begin{align}
E_{4}\left(\frac{\rho}{\kappa}\right)-E_{1}\left(\frac{\rho}{\kappa}\right)+\frac{\kappa}{\rho}E_{5}\left(\frac{\rho}{\kappa}\right)-\frac{\kappa}{\rho}E_{2}\left(\frac{\rho}{\kappa}\right)
=&\ \frac{1}{6}\left(\left(2-\frac{\rho}{\kappa}+\frac{\rho^{2}}{\kappa^{2}}\right)e^{-\frac{\rho}{\kappa}}-\frac{\rho^{3}}{\kappa^{3}}E_{1}\left(\frac{\rho}{\kappa}\right)\right) \nonumber \\
&-E_{1}\left(\frac{\rho}{\kappa}\right)-\frac{\kappa}{\rho}\left(e^{-\frac{\rho}{\kappa}}-\frac{\rho}{\kappa}E_{1}\left(\frac{\rho}{\kappa}\right)\right) \nonumber\\
&+\frac{\kappa}{24\rho}\left(\left(6-\frac{2\rho}{\kappa}+\frac{\rho^{2}}{\kappa^{2}}-\frac{\rho^{3}}{\kappa^{3}}\right)e^{-\frac{\rho}{\kappa}}+\frac{\rho^{4}}{\kappa^{4}}E_{1}\left(\frac{\rho}{\kappa}\right)\right) \nonumber\\
=&\ \frac{\kappa}{8\rho}\left(\frac{\rho^{3}}{\kappa^{3}}-\frac{\rho^{2}}{\kappa^{2}}+\frac{2\rho}{\kappa}-6\right)e^{-\frac{\rho}{\kappa}}-\frac{\rho^{3}}{8\kappa^{3}}E_{1}\left(\frac{\rho}{\kappa}\right).
\label{Ensim2}
\end{align}
Inserting Eqs. \ref{Ensim1} and \ref{Ensim2} into Eq. \ref{chi1GB} and adding in the general solution to Laplace's equation then yields
\begin{equation}
\begin{aligned}\label{eq:firstOrderInnerSol}
\chi_1 =&\ \frac{\gamma}{2} \sqrt{\frac{4\pi}{3}} Y_1^0\left[ \frac{\kappa^2}{4\rho^3}e^{1/\kappa}\left(\left( \frac{\rho^3}{\kappa^3} -\frac{\rho^2}{\kappa^2} +\frac{2\rho}{\kappa}-6\right)e^{-\rho/\kappa}-\frac{\rho^4}{\kappa^4}E_1\left( \frac{\rho}{\kappa}\right)\right) + 1 - \frac{2\kappa+1}{\rho^2} +\frac{1 + 3\kappa +3 \kappa^2}{2\rho^3}\right]\\
&+ \gamma \sum\limits_{\ell\geq 0}\left(A_{1,\ell}\rho^{\ell} +\frac{B_{1,\ell}}{\rho^{\ell+1}}\right)Y_{\ell}^0.
\end{aligned}
\end{equation}
It will be convenient to introduce a constant
\begin{equation}\label{eq:w}
    w = 1 + \kappa^{-1} + \kappa^{-2}e^{1/\kappa}E_1(\kappa^{-1}).
\end{equation}
The boundary condition from Eq.\ \ref{eq:firstOrderInnerPDE} translates to
\begin{equation}\label{eq:firstOrderInnerBC}
      \sqrt{\frac{4\pi}{3}} \frac{w}{8}\delta_{\ell,1} +\ell A_{1,\ell} -(\ell+1)B_{1,\ell} = \tilde{\alpha}\left[ \sqrt{\frac{4\pi}{3}}\frac{w}{8}\delta_{\ell,1} + A_{1,\ell} + B_{1,\ell}\right].
\end{equation}

\subsection{Asymptotic Matching}

We attempt to match the first-order solution for $\chi$ to the lowest order solution for $X$. We will see that this leads to a consistent matching condition, confirming that we may only work with the lowest order term in $X$.

As before, none of the terms in $X$ diverges at large $s$, so we need $A_{1,\ell}=0$ for $\ell\geq 1$. The boundary condition for the surface for $\chi_1$ in Eq.\ \ref{eq:firstOrderInnerBC} implies that $B_{1,\ell}=0$ for $\ell \geq 2$. Because $A_{1,1} = 0$, we set $\ell=1$ in Eq.\ \ref{eq:firstOrderInnerBC} to find
\begin{equation}\label{eq:chi1B11}
    B_{1,1} = \frac{w}{8} \sqrt{\frac{4\pi}{3}} \frac{1-\tilde{\alpha}}{2+\tilde{\alpha}}.
\end{equation}
We expand $X$ (Eq.\ \ref{0supp} with $s = \epsilon \rho$) to first order in $\epsilon$, giving
\begin{equation}
    X = \sqrt{4\pi} \frac{\gamma}{\rho} Y_0^0 +\frac{\epsilon \gamma}{2} \left(\sqrt{\frac{4\pi}{3}}Y_1^0 -\sqrt{4\pi} Y_0^0\right),
\end{equation}
where again we have used $Y_1^0 = \sqrt{4\pi/3}\cos\theta$.
Our form for $B_{1,1}$ is fine, since there is no term in $X$ proportional to $\rho^{-2}$ to this order and we are matching the large $\rho$ behavior of $\chi$ to $X$. Since the $\mathcal{O}(\epsilon^0)$ term in $X$ was matched by $\chi_0$, we must match $\epsilon \chi_1$ to the $\mathcal{O}(\epsilon)$ term in $X$. The $Y_1^0$ term in $X$ must be matched by the inhomogeneous term in $\chi_1$, as the terms in the homogeneous solution do not have a constant times $Y_1^0$. Specifically, we need the bracketed term in Eq.\ \ref{eq:firstOrderInnerSol} to tend to $1$ as $\rho\rightarrow \infty$. We have no parameters to tune, so if this fails, we will have to go to higher order. However, the limit is one, so this is consistent. We find $A_{1,0}$ from matching to the last term in $X$,
\begin{equation}
    A_{1,0} = -\sqrt{\pi }.
\end{equation}
Solving for $B_{1,0}$ using the boundary condition at the surface gives
\begin{equation}
    B_{1,0} = \frac{\tilde{\alpha} \sqrt{\pi}}{1+ \tilde{\alpha}}.
\end{equation}
These matching conditions are consistently satisfied, confirming that we may only work with the lowest order term in $X$. Using the values of the coefficients and the spherical harmonics to simplify Eq.\ \ref{eq:firstOrderInnerSol} gives
\begin{equation}
\label{chi1supp}
\chi_1 = \frac{\gamma}{2}\left\{\frac{\tilde{\alpha}}{(1+\tilde{\alpha})\rho}-1
	+\frac{\cos\theta}{4}\left[\frac{(1-\tilde{\alpha})w}{(2+\tilde{\alpha})\rho^2}
	+f(\rho,\kappa)\right]\right\},
\end{equation}
as in Eq.\ 5 of the main text, where the auxiliary function is
\begin{equation}
    f(\rho,\kappa) = 4 - \frac{4(2\kappa+1)}{\rho^2} + \frac{2(1+3\kappa+3\kappa^2)}{\rho^3}
    + \frac{\kappa^2e^{1/\kappa}}{\rho^3}\left[\left( \frac{\rho^3}{\kappa^3} - \frac{\rho^2}{\kappa^2} + \frac{2\rho}{\kappa} - 6\right)e^{-\rho/\kappa} - \frac{\rho^4E_1(\rho/\kappa)}{\kappa^4}\right].
\end{equation}
Note that $f(1,\kappa) = w$.

\section{Numerical validation of Eq.\ 5}

Here we verify that our perturbative solution is valid in the regime of interest for the P\'eclet number, $\epsilon \le 0.1$. Using Mathematica's \texttt{NDSolve} routine, we numerically solve the non-dimensionalized PDE
\begin{equation}
    0 = \nabla^2 \chi_{\text{num}}(\rho,\theta) - \epsilon \vec{u}(\rho,\theta)\cdot \nabla \chi_{\text{num}}(\rho,\theta)
\end{equation}
with non-dimensionalized Brinkman flow lines $\vec{u}(\rho,\theta)$ from Eq.\ \ref{eq:BrinkmannFlow}, subject to the boundary conditions at the cell surface $\rho = 1$ and outer radius $\rho = \rho_{\max}$
\begin{equation}
    \quad -\frac{\partial \chi_{\text{num}}(\rho,\theta)}{\partial \rho} \Big|_{\rho=1} = \tilde{\beta} - \tilde{\alpha} \chi_{\text{num}}(1,\theta), \quad \text{and} \quad  \chi_{\text{num}}(\rho_{\max}, \theta) = 0.
\end{equation}
The solution to this problem converges to the solution of our problem in the limit $\rho_{\max} \rightarrow \infty$.

\begin{figure}[h]
    \centering
    \includegraphics[width=0.9\textwidth]{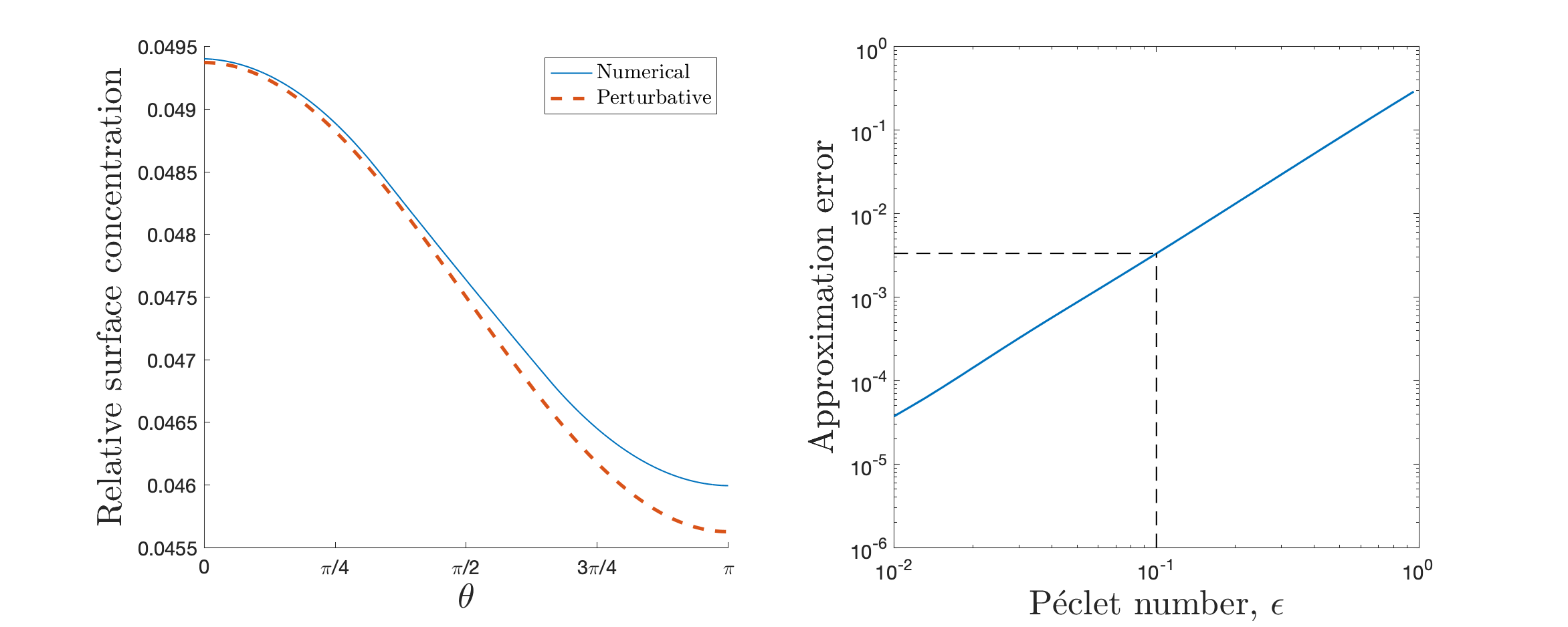}
    \caption{Left: Numerical (solid) and perturbative (dashed) solutions evaluated at the surface of the cell $\rho = 1$ for $\epsilon = 1$. Right: Approximation error between numerical and perturbative solutions (Eq.\ \ref{eq:apperr}). In both panels, $\tilde{\beta} = 0.05$, $\kappa = 10^{-3}$, $\alpha = 0$, and $\rho_{\max} = 10^3$.}
    \label{fig:pdeplot}
\end{figure}

We take $\tilde{\beta} = \beta a^4/D = 0.05$ and $\kappa = 10^{-3}$, corresponding to the typical experimental values listed in the main text, as well as $\alpha = 0$ and $\rho_{\max} = 10^3$. The left panel of Fig.\ \ref{fig:pdeplot} shows that for $\epsilon = 0.1$, the numerical solution $\chi_{\rm num}(1,\theta)$ and the perturbative solution $\chi_0(1) + \epsilon \chi_1(1,\theta)$ have close agreement at the surface of the cell.

We quantify the approximation error by
\begin{equation} \label{eq:apperr}
    \text{Error} = \frac{1}{4 \pi} \int\left| \frac{  \chi_{\text{num}}(1,\theta) - (\chi_0(1) + \epsilon \chi_1(1,\theta))  }{ \chi_{\text{num}}(1,\theta) }\right| \, d \Omega.
\end{equation}
We calculate the error for $100$ different $\epsilon$ values uniformly log-spaced between $10^{-2}$ and $1$. The results are shown in the right panel Fig.\ \ref{fig:pdeplot}. We see that the error is less than $0.4 \%$ when $\epsilon \leq 0.1$, a bound that encompasses our range of interest in $\epsilon$. We have checked that the results in Fig.\ \ref{fig:pdeplot} remain unchanged for $\rho_{\max} \gtrsim 750$, and therefore that our choice of $\rho_{\max} = 10^3$ is sufficiently large to avoid finite size effects.

\section{Derivation of Eq.\ 7}

The anisotropy in the absorption case (Eq.\ 6 of the main text) is
\begin{equation}
\label{Aasupp}
A \equiv \frac{\int_0^T dt \int a^2 d\Omega\ \alpha c(a,\theta,\phi,t) \cos\theta}{T \int a^2 d\Omega'\ \alpha \bar{c}(a,\theta')}.
\end{equation}
The mean of this expression is
\begin{equation}
\label{Aa2}
\bar{A} = \frac{\int d\Omega\ \bar{c}(a,\theta) \cos\theta}{\int d\Omega'\ \bar{c}(a,\theta')},
\end{equation}
where we have canceled the $T$, $\alpha$, and $a^2$. We evaluate these integrals using $\bar{c}(a,\theta) = [\chi_0(1,\theta) + \epsilon\chi_1(1,\theta)]/a^3$, where $\chi_0$ and $\chi_1$ are given by Eqs.\ \ref{0supp} and \ref{chi1supp}, respectively. In the numerator of Eq.\ \ref{Aa2}, the $\chi_0$ term vanishes because $\cos\theta$ integrates to zero. For the same reason, the only non-vanishing part of the $\chi_1$ term is the $\cos\theta$ term in Eq.\ \ref{chi1supp}, as the integral of $\cos^2\theta$ is nonzero. Here we also recall that $f(1,\kappa) = w$. In the denominator of Eq.\ \ref{Aa2}, the $\chi_0$ term is nonzero, and therefore we do not need the $\chi_1$ term to leading order. Altogether, Eq.\ \ref{Aa2} evaluates to
\begin{equation}
\label{meanasupp}
\bar{A} = \frac{w\epsilon}{8(2+\tilde{\alpha})},
\end{equation}
as in Eq.\ 7 of the main text.

The equivalent expression to Eq.\ \ref{Aasupp} that accounts for discrete molecule arrival, as stated in the main text, is
\begin{equation}
\label{Adiscrete}
A = \frac{1}{\bar{N}}\sum_{i=1}^N \cos\theta_i,
\end{equation}
where $\theta_i$ is the arrival angle of the $i$th molecule, and
\begin{equation}
\label{Ndef}
N = \int_0^T dt \int a^2d\Omega\ \alpha c(a,\theta,\phi,t)
\end{equation}
is the total number of molecules absorbed in time $T$. Here we will show that the mean of Eq.\ \ref{Adiscrete} also evaluates to Eq.\ \ref{meanasupp}. The mean of Eq.\ \ref{Adiscrete} is 
\begin{equation}
\label{Adis2}
\bar{A} =\frac{1}{\bar{N}} \left\langle \sum\limits_{i=1}^{N} \cos\theta_i \right\rangle,
\end{equation}
where the overbar and angle brackets are used interchangeably. Because the $N$ absorption events are statistically independent, the angle-bracketed term in Eq.\ \ref{Adis2} simply amounts to $\bar{N}$ copies of $\langle\cos\theta\rangle$. Thus,
\begin{equation}
\label{Adis3}
\bar{A} = \braket{\cos\theta}.
\end{equation}
The averaging is performed over the distribution defined by the mean surface concentration $\bar{c}(a,\theta)$. Explicitly,
\begin{equation}
\bar{A} = \frac{\int d\Omega\ \bar{c}(a,\theta) \cos\theta}{\int d\Omega'\ \bar{c}(a,\theta')}.
\end{equation}
This expression is equivalent to Eq.\ \ref{Aa2} and therefore evaluates to Eq.\ \ref{meanasupp}.

Note that the definition of $A$ implicitly assumes that the cell ``knows'' the true direction of the flow to be $\theta = 0$. In reality this is untrue. Instead, the migration direction of the cell is a three-dimensional vector that can be decomposed into three components along the $\hat{x}$, $\hat{y}$, and $\hat{z}$ ($\theta = 0$) directions. However, the means of the components in the $\hat{x}$ and $\hat{y}$ directions involve
averages of $\sin\theta \cos\phi$ and $\sin\theta \sin\phi$, which are zero due to the azimuthal symmetry. Therefore, the result in Eq.\ 7 holds even when accounting for all three components.

\section{Derivation of Eq.\ 8}

To compute the variance of Eq.\ \ref{Adiscrete}, we use the fact that the number of molecules absorbed in a patch on the cell surface is a Poisson variable (confirmed with simulations in the next section). Letting $\theta_i$ denote the value of $\theta$ at which particle $i$ is absorbed, the second moment of the sum of cosines is
\begin{equation} \label{eq:2mom}
    \left\langle \left( \sum\limits_{i=1}^{N} \cos\theta_i \right)^2 \right\rangle = \left\langle \sum\limits_{i=1}^{N} \cos^2\theta_i \right\rangle +\left\langle \sum\limits_{i\neq j} \cos\theta_i \cos\theta_j  \right\rangle
    = \braket{N} \braket{\cos^2\theta} +\braket{N(N-1)} \braket{\cos\theta}^2,
\end{equation}
where again the second step follows from the fact that the absorption events are statistically independent. For a Poisson random variable
\begin{equation} \label{Poissrand}
    \braket{N} =\sigma_N^2 = \braket{N^2} -\braket{N}^2 \Longrightarrow \braket{N(N-1)} = \braket{N}^2.
\end{equation}
Inserting this result into Eq.\ \ref{eq:2mom}, we see that the last term becomes the square of the mean and will thus cancel when using Eq.\ \ref{eq:2mom} to calculate the variance. Additionally, we will need to multiply the variance by a factor of three. The reason is that $\cos^2\theta$ is an even function, and therefore the angular average, to lowest order in $\epsilon$, will be over only the uniform part of the solution ($\chi_0$). It will therefore have the same contributions from the $\hat{x}$ and $\hat{y}$ directions. Altogether, this allows us to write the variance as
\begin{equation} \label{var3}
    \sigma_A^2 = \frac{3}{\bar{N}^2} \text{Var}\left( \sum\limits_{i=1}^{N} \cos\theta_i \right) = \frac{3}{\bar{N}^2}\bar{N} \braket{\cos^2\theta} = \frac{3}{\bar{N}} \braket{\cos^2\theta}.
\end{equation}
The leading order terms in the averages of both $N$ (Eq.\ \ref{Ndef}) and $\cos^2\theta$ come only from the uniform $\chi_0$ (Eq.\ \ref{0supp}). Specifically,
\begin{equation}
    \bar{N} = \int_0^T dt \int a^2d\Omega\ \alpha \bar{c}(a,\theta) = a^2\alpha T \int d\Omega \frac{\gamma}{a^3} = \frac{4\pi \alpha \gamma T}{a} = \frac{\nu T\tilde{\alpha}}{1+\tilde{\alpha}},
\end{equation}
and the average of $\cos^2 \theta$ over the uniform sphere is
\begin{equation}
    \braket{\cos^2\theta} = \frac{1}{4\pi}\int d\Omega\ \cos^2\theta = \frac{1}{3}.
\end{equation}
Together these results produce Eq.\ 8 in the main text.

\section{Verification of Poisson statistics with particle-based simulations}

To verify the assumption that molecule absorption events at the surface of the cell follow a Poisson distribution, we use particle-based simulation. After non-dimensionalizing the problem and Brinkman flow equations, particles are pseudorandomly initialized between two spherical boundaries at $\rho = 1$ and $\rho = \rho_{\max} = 10$. Throughout the simulation, particles are generated at a random position on the cell surface with rate $\tilde{\beta} = 10$ (this value is larger than that estimated from experiments in order to generate good statistics in a reasonable computational time). Diffusion is discrete in time and continuous in space: in a dimensionless time step $\Delta\tau$, for each particle, we draw three samples from a normal distribution with mean zero and variance $2\Delta\tau$ (corresponding to a variance of $2D\Delta t$ in real units) for each spatial component. The absorption propensity $\tilde{\alpha} = 0.75$ is used to determine absorption or reflection events for particles found within $\rho < 1 + \ell/a$, with $\ell/a = 0.01$ interpreted as a maximal receptor height. For recording, the cell surface is split into 100 ring-shaped patches over $\theta \in [0, \pi]$, uniform in $\cos \theta$. Particles are deleted whenever they diffuse past the outer boundary.

The four timescales in the system are the birth, diffusive, drift, and absorption timescales, which are $1/4 \pi a^2 \beta$, $(a\Delta \rho)^2/D$, $a/v_0$, and $\ell/\alpha$, respectively, where $\Delta \rho = (\rho_{\max}-1)/100$. In dimensionless units these timescales read $1/4 \pi \tilde\beta$, $(\Delta \rho)^2$, $1/\epsilon$, and $\ell/\tilde{\alpha}a$, respectively. The time step is set to be smaller than all four timescales, at $\Delta\tau = 0.001$. The simulation is run for $2\times10^5$ time steps.

\begin{figure}[h]
\centering
	\includegraphics[width=\textwidth]{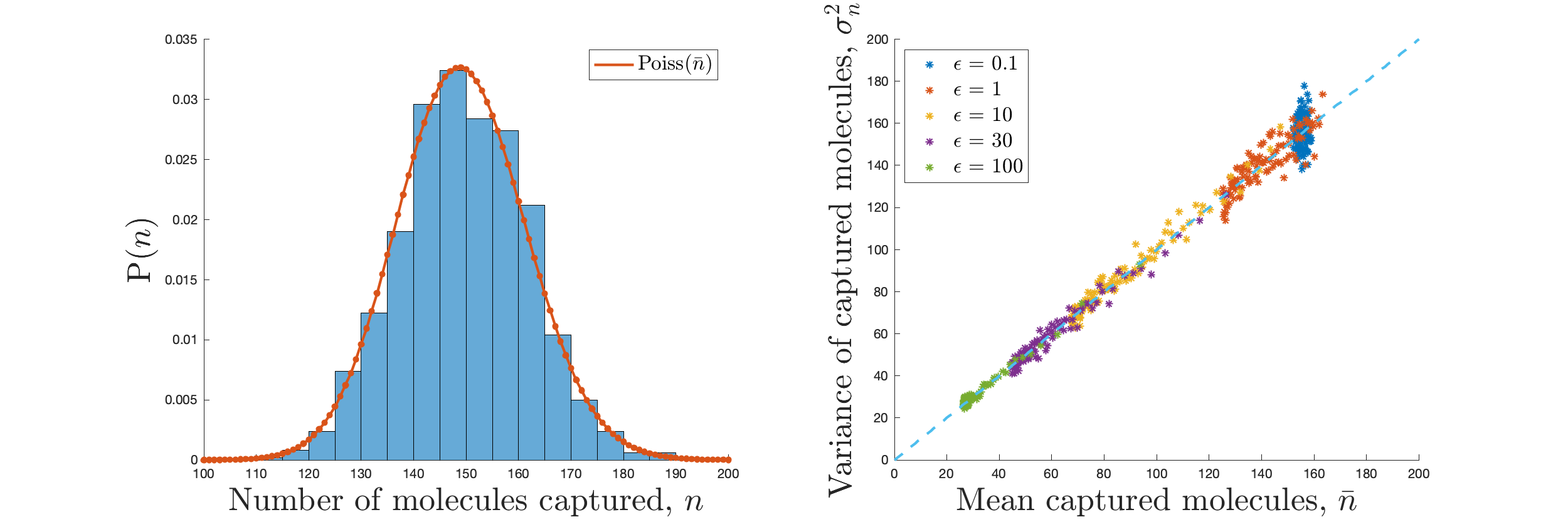}
	\caption{ Left: An example of the distribution of $n$ over 1000 trials at a particular patch ($\theta = \pi/3$) for P\'eclet number $\epsilon = 1$, plotted against a Poisson distribution with equal sample mean. Right: The sample variances $\sigma_n^2$ from all patches of the cell (data points) lie close to the sample mean $\overline{n}$ for multiple values of the P\'eclet number $\epsilon$.}
	\label{sim}
\end{figure}

The number of molecules $n$ absorbed at a particular patch ($\theta = \pi/3$) with $\epsilon = 1$ across an ensemble of 1000 trials is shown in Fig.\ \ref{sim} (left). We see that the distribution of $n$ is in excellent agreement with a Poisson distribution of the same mean. We repeat this measurement for all patches and with different values of the P\'eclet number $\epsilon$ in Fig.\ \ref{sim} (right). We see that, consistent with Poisson statistics, the data fall along the line for which the variance equals the mean, even up to a P\'eclet number of $\epsilon = 100$.

\section{Derivation of Eq.\ 14}

As in the absorption case, we non-dimensionalize the system in Eq.\ 11 of the main text for the binding case. We use the same parameters from Eq.\ \ref{eq:nondimabs} where relevant and introduce the new parameters
\begin{equation}\label{eq:nondimrev}
    \begin{gathered}
        \psi= a^2 b, \qquad \tilde{\lambda} = \frac{\lambda a}{D}, \qquad \tilde{\mu} = \frac{\mu a^2}{D}, \qquad \tau = \frac{t D}{a^2}, \\
        \xi_D = \frac{a^5}{D} \eta_D, \qquad \xi_{\beta} = \frac{a^4}{D} \eta_{\beta}, \qquad \xi_{b} = \frac{a^4}{D} \eta_{b}.
    \end{gathered}
\end{equation}
The non-dimensionalized versions of Eqs.\ 11 and 12 of the main text are then
\begin{equation}\label{eq:revSPDEnondim}
    \begin{aligned}
        \frac{\partial \chi}{\partial \tau}&=\nabla_{\rho}^{2} \chi-\epsilon \vec{u} \cdot \vec{\nabla}_{\rho} \chi+\xi_D+\left(-\frac{\partial \psi}{\partial \tau}+\tilde{\beta}+\xi_\beta\right) \delta(\rho-1) \\
        \frac{\partial \psi}{\partial \tau}&=\tilde{\lambda} \chi(1,\hat{\Omega})-\tilde{\mu} \psi+\xi_b.
    \end{aligned}
\end{equation}
and
\begin{equation}\label{eq:nondimcovar}
    \begin{gathered}
    \left\langle\xi_D\left(\vec{\rho}\ ', \tau^{\prime}\right) \xi_D(\vec{\rho}, \tau)\right\rangle= 2 \delta\left(\tau-\tau^{\prime}\right) \vec{\nabla}_{\rho} \cdot \vec{\nabla}_{\rho^{\prime}}\left[\bar{\chi}(\vec{\rho}) \delta^{3}(\vec{\rho}-\vec{\rho}\text{ }')\right],\\
    \left\langle\xi_\beta\left(\hat{\Omega}^{\prime}, \tau^{\prime}\right) \xi_\beta(\hat{\Omega}, \tau)\right\rangle=\tilde{\beta} \delta\left(\tau-\tau^{\prime}\right) \delta^{2}\left(\hat{\Omega}-\hat{\Omega}^{\prime}\right), \\
    \left\langle\xi_b\left(\hat{\Omega}^{\prime}, \tau^{\prime}\right) \xi_b(\hat{\Omega}, \tau)\right\rangle= 2 \tilde{\mu} \bar{\psi} \delta\left(\tau-\tau^{\prime}\right) \delta^{2}\left(\hat{\Omega}-\hat{\Omega}^{\prime}\right),
    \end{gathered}
\end{equation}
where we use $\hat{\Omega}$ to denote the solid angle $(\theta, \phi)$. In general below, we will use a hat to denote the angular components of a vector.

We will find the variance in the signal by using the Wiener-Khinchin theorem and noting that the zero frequency limit of the power spectrum gives the long time behavior of the variance. We start by linearizing Eq.\ \ref{eq:revSPDEnondim} using
\begin{equation}
    \delta \chi = \chi -\overline{\chi}, \qquad \delta \psi = \psi - \overline{\psi}.
\end{equation}
We then Fourier transform $\delta\chi$ and $\delta\psi$ as
\begin{equation}\label{eq:fourierT}
\begin{gathered}
\tilde{\delta\chi}\left(\vec{k},\omega\right) = \int d^{3}\rho d\tau\ e^{i\vec{k}\cdot\vec{\rho}}e^{i\omega\tau}\delta\chi\left(\vec{\rho},\tau\right), \\
\tilde{\delta\psi}_{\ell}^{m}\left(\omega\right) = \int d\Omega d\tau\ Y_{\ell}^{m}\left(\hat{\Omega}\right)e^{i\omega\tau}\delta\psi\left(\hat{\Omega},\tau\right).
\end{gathered}
\end{equation}
Linearizing and transforming $\chi$ and $\psi$ in these ways have two important effects on Eq.\ \ref{eq:revSPDEnondim}. First, the linearization eliminates the $\tilde{\beta}$ term, and second, the Fourier transformation allows for derivatives with respect to $\vec{\rho}$ and $\tau$ to be written in Fourier space as $-i\vec{k}$ and $-i\omega$ respectively. In addition, we make the approximation $\epsilon=0$ as that is the lowest order term in the variance of each dynamic variable. These effects transform Eq.\ \ref{eq:revSPDEnondim} into
\begin{align}
-i\omega\tilde{\delta\chi} &= -k^{2}\tilde{\delta\chi}+\tilde{\xi}_{D}+\int d^{3}\rho\> e^{i\vec{k}\cdot\vec{\rho}}\delta\left(\rho-1\right)\sum_{\ell,m}Y_{\ell}^{m*}\left(\hat{\rho}\right)\left(i\omega\tilde{\delta\psi}_{\ell}^{m}+\tilde{\xi}_{\beta\ell}^m\right) \nonumber\\
&= -k^{2}\tilde{\delta\chi}+\tilde{\xi}_{D}+4\pi\sum_{\ell,m}i^{\ell}j_{\ell}\left(k\right)Y_{\ell}^{m*}\left(\hat{k}\right)\left(i\omega\tilde{\delta\psi}_{\ell}^{m}+\tilde{\xi}_{\beta\ell}^m\right),
\label{LFTnondiffdrifprodeq}
\end{align}
and
\begin{align}
-i\omega\tilde{\delta\psi}_{\ell}^{m} &= \tilde{\lambda}\int d\Omega\>Y_{\ell}^{m}\left(\hat{\Omega}\right)\int\frac{d^{3}k}{\left(2\pi\right)^{3}}e^{-i\hat{\Omega}\cdot\vec{k}}\tilde{\delta\chi}\left(\vec{k},\omega\right)-\tilde{\mu}\tilde{\delta\psi}_{\ell}^{m}+\tilde{\xi}_{b\ell}^m \nonumber\\
&= 4\pi\tilde{\lambda}\int\frac{d^{3}k}{\left(2\pi\right)^{3}}\left(-i\right)^{\ell}j_{\ell}\left(k\right)Y_{\ell}^{m}\left(\hat{k}\right)\tilde{\delta\chi}\left(\vec{k},\omega\right)-\tilde{\mu}\tilde{\delta\psi}_{\ell}^{m}+\tilde{\xi}_{b\ell}^m,
\label{LFTnonbindeq}
\end{align}
where Eqs.\ \ref{LFTnondiffdrifprodeq} and \ref{LFTnonbindeq} have been simplified using the plane wave expansion
\begin{equation}
e^{i\vec{x}\cdot\vec{y}} = 4\pi\sum_{\ell,m}i^{\ell}j_{\ell}\left(xy\right)Y_{\ell}^{m}\left(\hat{x}\right)Y_{\ell}^{m*}\left(\hat{y}\right),
\label{planewaveexp}
\end{equation}
and $j_\ell$ denotes a spherical Bessel function of the first kind.

From here we set $\omega=0$ to obtain the long time dynamics. Making this substitution along with solving Eq.\ \ref{LFTnondiffdrifprodeq} for $\tilde{\delta\chi}$ and substituting that solution into Eq.\ \ref{LFTnonbindeq} produces the relation
\begin{equation}
0 = 4\pi\left(-i\right)^{\ell}\tilde{\lambda}\int\frac{d^{3}k}{\left(2\pi\right)^{3}}\frac{1}{k^{2}}j_{\ell}\left(k\right)Y_{\ell}^{m}\left(\hat{k}\right)\tilde{\xi}_{D}+\frac{\tilde{\lambda}}{2\ell+1}\tilde{\xi}_{\beta\ell}^m-\tilde{\mu}\tilde{\delta\psi}_{\ell}^{m}+\tilde{\xi}_{b\ell}^m,
\label{eq:LFTpsirel}
\end{equation}
where the orthonormality of spherical harmonics and the known properties of spherical Bessel functions have been used to simplify the result. Solving Eq.\ \ref{eq:LFTpsirel} for $\tilde{\delta\psi}_{\ell}^{m}$ and using the fact that each $\xi$ term is independent of the others then yields
\begin{align}\label{psicross}
    \left\langle\tilde{\delta\psi}_{\ell'}^{m'*}\tilde{\delta\psi}_{\ell}^{m}\right\rangle =&\ \frac{{\tilde{\lambda}}^{2}}{{\tilde{\mu}}^{2}\left(2\ell+1\right)\left(2\ell'+1\right)}\left\langle\tilde{\xi}_{\beta\ell'}^{m'*}\tilde{\xi}_{\beta\ell}^m\right\rangle+\frac{1}{{\tilde{\mu}}^{2}}\left\langle\tilde{\xi}_{b\ell'}^{m'*}\tilde{\xi}_{b\ell}^m\right\rangle \nonumber \\
&+\frac{i^{\ell'-\ell}{\tilde{\lambda}}^{2}}{4\pi^{4}{\tilde{\mu}}^{2}}\int d^{3}kd^{3}k'\frac{j_{\ell}\left(k\right)j_{\ell'}\left(k'\right)}{k^{2}{k'}^{2}}Y_{\ell}^{m}\left(\hat{k}\right)Y_{\ell'}^{m'*}\left(\hat{k}'\right)\left\langle{\tilde{\xi}_{D}}^{*}\left(\vec{k}'\right)\tilde{\xi}_{D}\left(\vec{k}\right)\right\rangle.
\end{align}
Taking the Fourier transform for the diffusive noise covariance in Eq.\ \ref{eq:nondimcovar} and integrating the gradient terms by parts gives
\begin{equation}
\left\langle{\tilde{\xi}_{D}}^{*}\left(\vec{k}',\omega'\right)\tilde{\xi}_{D}\left(\vec{k},\omega\right)\right\rangle = 2\left(2\pi\delta\left(\omega-\omega'\right)\right)\vec{k}\cdot\vec{k}'\int d^{3}\rho\>\bar{\chi}\left(\vec{\rho}\right)e^{i\vec{\rho}\cdot\left(\vec{k}-\vec{k}'\right)}.
\label{etaDcross}
\end{equation}
The covariances for the binding-unbinding and production reactions are
\begin{equation}\label{etabetacross}
\begin{gathered}
\left\langle\tilde{\xi}_{\beta\ell'}^{m'*}\left(\omega'\right)\tilde{\xi}_{\beta\ell}^m\left(\omega\right)\right\rangle =  \tilde{\beta}\delta_{\ell\ell'}\delta_{mm'}\left(2\pi\delta\left(\omega-\omega'\right)\right), \\
\left\langle\tilde{\xi}_{b\ell'}^{m'*}\left(\omega'\right)\tilde{\xi}_{b\ell}^m\left(\omega\right)\right\rangle = 2\tilde{\beta}\tilde{\lambda}\delta_{\ell\ell'}\delta_{mm'}\left(2\pi\delta\left(\omega-\omega'\right)\right).
\end{gathered}
\end{equation}
where the $\epsilon=0$ approximation with $\alpha = 0$ has been applied to Eq.\ \ref{0supp} to write $\bar{\psi} = \tilde{\lambda}\bar{\chi}_0(1,\hat{\Omega})/\tilde{\mu} = \tilde{\lambda}\gamma/\tilde{\mu} = \tilde{\lambda}\tilde{\beta}/\tilde{\mu}$.

If $x$ is some real, stationary, ergodic process, we define its power spectrum through
\begin{equation}
    \braket{\tilde{x}(\omega) \tilde{x}^*(\omega')} = 2\pi S(\omega) \delta(\omega-\omega'),
\end{equation}
where $\tilde{x}(\omega)$ is the Fourier transform of $x(t)$ using the same sign and normalization convention as in Eq.\ \ref{eq:fourierT}. The long time behavior of the variance in the time average of $x(t)$ is 
\begin{equation} \label{STrel}
    \sigma^2(T) = \frac{S(0)}{T}.
\end{equation}
Eqs.\ \ref{etaDcross} and \ref{etabetacross} all have factors of $2\pi\delta(\omega-\omega')$, so we can obtain the power spectrum for the components of $\psi$ by neglecting these factors. Doing so while inserting Eqs.\ \ref{etaDcross} and \ref{etabetacross} into Eq.\ \ref{psicross} then yields the cross-spectrum between the $(\ell,m)$ and $(\ell', m')$ components of $\psi$ at $\omega=0$
\begin{equation} 
S_{\ell\ell'mm'}^{(\psi)}(0) = \frac{\tilde{\beta}{\tilde{\lambda}}^{2}}{{\tilde{\mu}}^{2}}\left(i^{\ell'-\ell}I_{\ell\ell'mm'}+\left(\frac{1}{\left(2\ell+1\right)\left(2\ell'+1\right)}+\frac{2}{\tilde{\lambda}}\right)\delta_{\ell\ell'}\delta_{mm'}\right),
\label{psicross2}
\end{equation}
where
\begin{equation}
I_{\ell\ell'mm'} = \frac{1}{2\pi^{4}\tilde{\beta}}\int d^{3}kd^{3}k'd^{3}\rho\frac{j_{\ell}\left(k\right)j_{\ell'}\left(k'\right)}{k^{2}{k'}^{2}}Y_{\ell}^{m}\left(\hat{k}\right)Y_{\ell'}^{m'*}\left(\hat{k}'\right)\vec{k}\cdot\vec{k}'\bar{\chi}\left(\vec{\rho}\right)e^{i\vec{\rho}\cdot\left(\vec{k}-\vec{k}'\right)}.
\label{Illmmdef}
\end{equation}

From here, $I_{\ell\ell'mm'}$ needs to be simplified. To do so, we will first apply the $\epsilon=0$ approximation and $\alpha = 0$ to Eq.\ \ref{0supp} to write $\bar{\chi} = \tilde{\beta}/\rho$. This will also restrict the $\rho$ integral to be only over the space where $\rho\ge 1$ as molecules are considered to not exist inside the cell. Additionally, the exponential piece can be expanded via Eq.\ \ref{planewaveexp}. Eq.\ \ref{eq:legendre2Spherical} can be used to simplify $\hat{k}\cdot \hat{k}' = P_1(\hat{k}\cdot \hat{k}')$. Performing these expansions and using orthogonality of spherical harmonics allows $I_{\ell\ell'mm'}$ to be simplified into
\begin{align}
&I_{\ell\ell'mm'} = \frac{16\sqrt{\left(2\ell+1\right)\left(2\ell'+1\right)}}{\pi^{2}} \sum\limits_{\ell''=0}^{\infty} \sum\limits_{m''=-\ell''}^{\ell''} \sum_{m'''=-1}^{1} \int_{1}^{\infty}d\rho\int_{0}^{\infty}dkdk' \nonumber\\
&\quad\quad\quad\quad \times \left(2\ell''+1\right)\rho kk'j_{\ell}\left(k\right)j_{\ell'}\left(k'\right)j_{\ell''}\left(\rho k\right)j_{\ell''}\left(\rho k'\right) \label{Illmmsimp} \\
&\quad\quad\quad\quad \times\begin{pmatrix}
\ell & 1 & \ell'' \\
0 & 0 & 0 \end{pmatrix}\begin{pmatrix}
\ell & 1 & \ell'' \\
m & m''' & m'' \end{pmatrix}\begin{pmatrix}
\ell' & 1 & \ell'' \\
0 & 0 & 0 \end{pmatrix}\begin{pmatrix}
\ell' & 1 & \ell'' \\
m' & m''' & m'' \end{pmatrix}. \nonumber
\end{align}
where the final line contains Wigner 3-j symbols.

Before continuing to simplify $I_{\ell\ell'mm'}$, we first consider $A$ (Eq.\ 12 in the main text),
\begin{equation}
\label{Absupp}
A \equiv \frac{\int_0^T dt \int a^2d\Omega\ b(\theta,\phi,t)\cos\theta}{T \int a^2 d\Omega'\ \bar{b}(\theta')}.
\end{equation}
We write this equation as $A = T^{-1}\int_0^T dt\ B(t)$, where
\begin{equation}
B \equiv \frac{\int a^2d\Omega\ b(\theta,\phi,t)\cos\theta}{\int a^2 d\Omega'\ \bar{b}(\theta')}.
\end{equation}
Here, $A$ is the time average of $B$. Therefore, from Eq.\ \ref{STrel}, the variance of $A$  in the long-time limit will be given by the power spectrum of $B$ at zero frequency divided by the integration time $T$. We apply a similar series of linearization and Fourier transformation along with the approximation $\epsilon=0$ to $B$ to yield
\begin{equation}
\tilde{\delta B}\left(\omega\right) = \frac{a^2}{D}\int d\tau\> e^{i\omega\tau}\frac{\int d\Omega\>\cos\left(\theta\right)\delta\psi\left(\hat{\Omega},\tau\right)}{\int d\Omega'\>\bar{\psi}\left(\theta'\right)}
= \frac{\tilde{\mu}a^2}{\sqrt{12\pi}\tilde{\lambda}\tilde{\beta}D}\tilde{\delta\psi}_{1}^{0}\left(\omega\right).
\label{LFTC}
\end{equation}
Given that $\tilde{\delta B}$ depends only on $\tilde{\delta\psi}_{1}^{0}$, this is the only moment we need to solve for. Due to the selection rules of Wigner 3-j symbols, in the case where $\ell=\ell'=1$ and $m=m'=0$ the summations in Eq.\ \ref{Illmmsimp} only have four surviving terms: the $\ell''=m''=m'''=0$ term and the three $\ell''=2$, $m''=-m'''$ terms. These terms can be solved individually using the relations
\begin{equation}
\begin{gathered}
\int_{0}^{\infty}dk\>kj_{1}\left(k\right)j_{0}\left(\rho k\right) = 0, \\
\int_{1}^{\infty}d\rho\>\rho\left(\int_{0}^{\infty}dk\> kj_{1}\left(k\right)j_{2}\left(\rho k\right)\right)^{2} = \int_{1}^{\infty}d\rho\frac{\pi^{2}}{4\rho^{5}} = \frac{\pi^{2}}{16}, \\
\sum_{m'''=-1}^{1}\begin{pmatrix}
1 & 1 & 2 \\
0 & 0 & 0 \end{pmatrix}^{2}\begin{pmatrix}
1 & 1 & 2 \\
0 & m''' & -m''' \end{pmatrix}^{2} = \frac{2}{45}.
\end{gathered}
\end{equation}
Inserting these values and $\ell=\ell'=1$ and $m=m'=0$ into Eq.\ \ref{Illmmsimp} yields
\begin{equation}
I_{1100} = \frac{48}{\pi^{2}}\cdot 5\cdot\frac{\pi^{2}}{16}\cdot\frac{2}{45} = \frac{2}{3}.
\label{I1100}
\end{equation}
Inserting this into Eq.\ \ref{psicross2} then yields
\begin{equation}
S_{1100}^{(\psi)}(0) = \frac{\tilde{\beta}{\tilde{\lambda}}^{2}}{{\tilde{\mu}}^{2}}\left(\frac{7}{9}+\frac{2}{\tilde{\lambda}}\right),
\label{psicross1100}
\end{equation}
which when combined with Eq.\ \ref{LFTC} produces
\begin{equation}
S^{(B)}(0) = \frac{{\tilde{\mu}}^{2}a^2}{12\pi{\tilde{\beta}}^{2}{\tilde{\lambda}}^{2}D}S_{1100}^{(\psi)}(0) = \frac{a^2}{12\pi\tilde{\beta}D}\left(\frac{7}{9}+\frac{2}{\tilde{\lambda}}\right).
\label{Ccross}
\end{equation}
Finally, we use Eq.\ \ref{STrel} to write
\begin{equation}
\sigma_{A}^{2} = \frac{3S^{(B)}(0)}{T} = \frac{a^2}{4\pi\tilde{\beta}DT}\left(\frac{7}{9}+\frac{2}{\tilde{\lambda}}\right) = 
\frac{1}{\nu T}\left(\frac{7}{9}+\frac{2}{\tilde{\lambda}}\right),
\label{LTvarC}
\end{equation}
as in Eq.\ 14 of the main text, where the factor of $3$ accounts for the equivalent variance in the $\sin\theta\cos\phi$ and $\sin\theta\sin\phi$ components, and the last step recalls $\tilde{\beta} = \beta a^4/D$ and $\nu = 4\pi a^2\beta$.

\section{Effect of receptor clustering}

To test the robustness of the sensing mechanism to a nonuniform distribution of receptors on the surface of the cell, we performed a numerical analysis for the absorbing model. We consider the convection-diffusion equation, Eq.\ 2 in the main text, subject to a different boundary condition at the surface of the cell,
\begin{equation}
-\left.D \frac{\partial \bar{c}(r, \theta, \phi)}{\partial r}\right|_{r = a}=\beta-\alpha^{m}_{\ell} (\theta, \phi) \bar{c}(a, \theta, \phi),
\end{equation}
where the absorption rate $\alpha^{m}_{\ell}(\theta, \phi)$ is no longer a constant, but a function proportional to the real part of the spherical harmonic $Y^{m}_{\ell}(\theta, \phi)$ defined on the surface of the cell. Along with the condition that the concentration $\bar{c}(r, \theta, \phi)$ vanishes for large $r$, these equations constitute a well-posed problem that describes the case in which receptors ``cluster" on the surface of the cell. This clustering phenomenon is parametrized by the spherical harmonic numbers $\ell$ and $m$.
	
\begin{figure}[h]
\includegraphics[width=\textwidth]{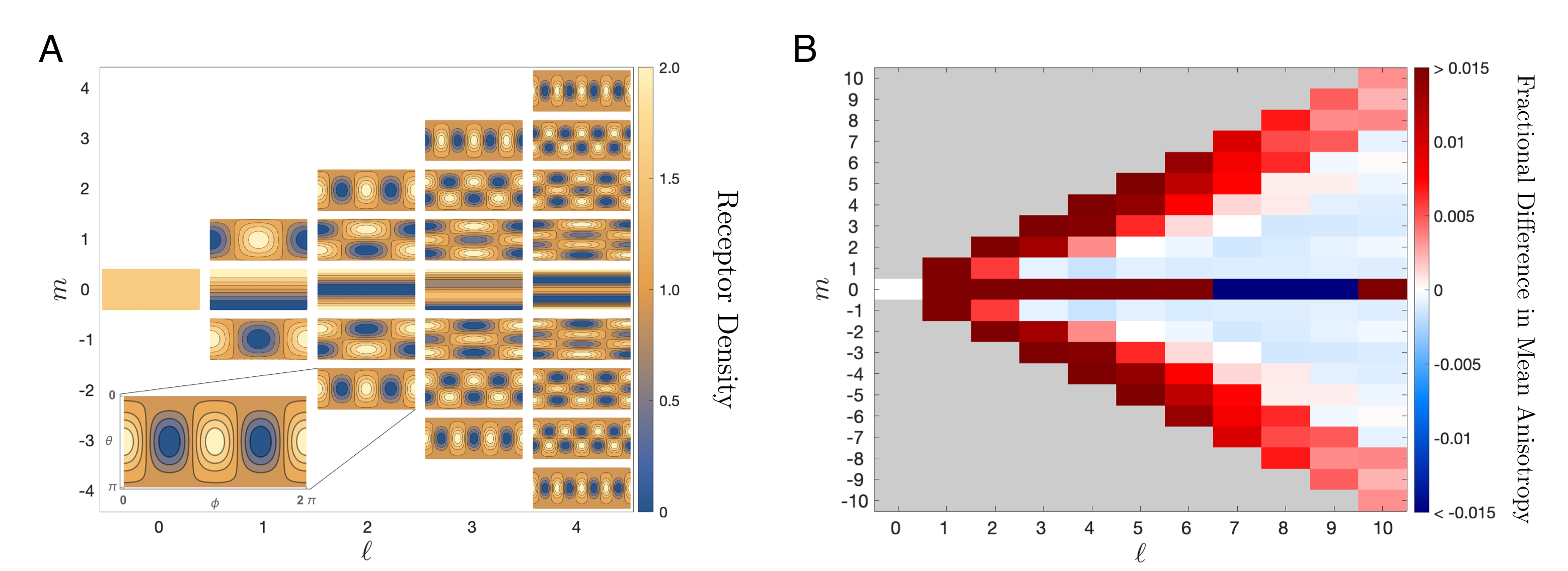}
\caption{Results are robust to receptor clustering. (A) Normalized dimensionless absorption rate $\tilde{\alpha}^{m}_{\ell}/\tilde{\alpha}$ as a function of $\theta$ and $\phi$ (see inset), constructed from spherical harmonics, which models an inhomogeneous receptor density. (B) Signed, fractional difference in average anisotropy measure. We see that difference is less than a percent for the ``patchy'' receptor configurations. Here $\tilde{\alpha} = \tilde{\alpha}^* = (\sqrt{17}-1)/4 \approx 0.78$, $\tilde{\beta} = 0.05$, $\epsilon = 0.01$, and $\rho_{\max} = 100$.}
\label{figS3}
\end{figure}

Specifically, after non-dimensionalizing the problem as in the main text, we define the dimensionless absorption rate
\begin{equation}
\tilde{\alpha}^{m}_{\ell}(\theta, \phi) = 2 \tilde{\alpha} \left( \frac{\Re \left[ Y^{m}_{\ell}(\theta, \phi) \right] - \min_{\theta, \phi} \Re \left[ Y^{m}_{\ell}(\theta, \phi) \right]}{ \max_{\theta, \phi} \Re \left[ Y^{m}_{\ell}(\theta, \phi) \right] - \min_{\theta, \phi} \Re \left[ Y^{m}_{\ell}(\theta, \phi) \right]} \right) \quad \text{for $\ell > 0$},
\end{equation}
where $\tilde{\alpha}$ is the constant dimensionless absorption rate used in the main text, equivalent to the $\ell = m = 0$ case here. This choice of $\tilde{\alpha}^{m}_\ell$ guarantees non-negativity as well as a spatial average equal to $\tilde{\alpha}$ for any choice of $\ell$ and $m$. Fig.\ \ref{figS3}A shows plots of $\tilde{\alpha}^{m}_\ell/\tilde{\alpha}$ over the cell surface for various choices of $\ell$ and $m$. We see that receptors are arranged in striped patterns for $m = 0$ and $|m| = \ell$ (the ``backbone'' and ``sides'' of the triangle) but that receptors are arranged in patches for other values of $m$ and $\ell$ (the ``bulk'' of the triangle). The latter case is more relevant to receptor clustering.
	
We solve the convection-diffusion equation for the ligand concentration $\bar{c}$ using Mathematica's \texttt{NDSolve} routine as in Section III. We evaluate the mean anisotropy measure $\bar{A}^{m}_{\ell}$ (Eq.\ 6 of the main text) from the absorptive flux $\tilde{\alpha}^{m}_{\ell} \bar{c}$. We compute a signed, normalized difference measure from the uniform case ($\ell = m = 0$),
\begin{align*}
 \text{Fractional difference in mean anisotropy} = \frac{\bar{A}^{m}_{\ell} - \bar{A}_0^0 }{\bar{A}_0^0}.
\end{align*}
Fig.\ \ref{figS3}B shows this difference measure as a function of $\ell$ and $m$. We see that in the bulk of the triangle, the measure shows a clear convergence toward zero as $\ell$ and $m$ become large. In fact, we see that for $\ell \gtrsim 5$, the difference is less than a percent. Receptor clusters usually occur on lengthscales much smaller than the cell size, corresponding to $\ell$ and $|m|$ values much larger than those shown. We conclude that receptor clustering has a negligible effect on our results.

\section{Effect of non-spherical cell geometry}

Cells polarize and stretch in the direction of motion as they move. In this section, we investigate the effect that stretching has on the ability of the cell to sense the direction of the fluid flow. For simplicity, we incorporate stretching (or compressing) by investigating an ellipsoidal cell. We also ignore the effect of impermeability of the medium and simply use Stokes' flow. For a spherical cell, we find in the main text that the impermeability halves the error (taking $w$ from 1 to 2) but does not change the overall scaling, and we expect the effect to be similar here. We also focus only on the absorbing case, where the deterministic convection-diffusion equation suffices to determine the statistics of the anisotropy measure.

\subsection{Ellipsoidal coordinate system}

We will find it useful to adapt our coordinates to the shape of the cell surface in order to state the boundary condition for the convection-diffusion equation. Therefore, we first introduce a new coordinate system that we will call ``ellipsoidal" for simplicity. Our new coordinates are not the same as the standard confocal ellipsoidal coordinates or the prolate or oblate spheroidal coordinates. They are also non-orthogonal, and therefore they give rise to off-diagonal terms in definitions such as the Laplacian, as we derive using differential geometry in a later section below. Nonetheless, they are a continuous deformation of spherical coordinates and are useful for specifying the boundary and visualizing the system.

\begin{figure}[b]
	\begin{center}
		\includegraphics[width=\textwidth]{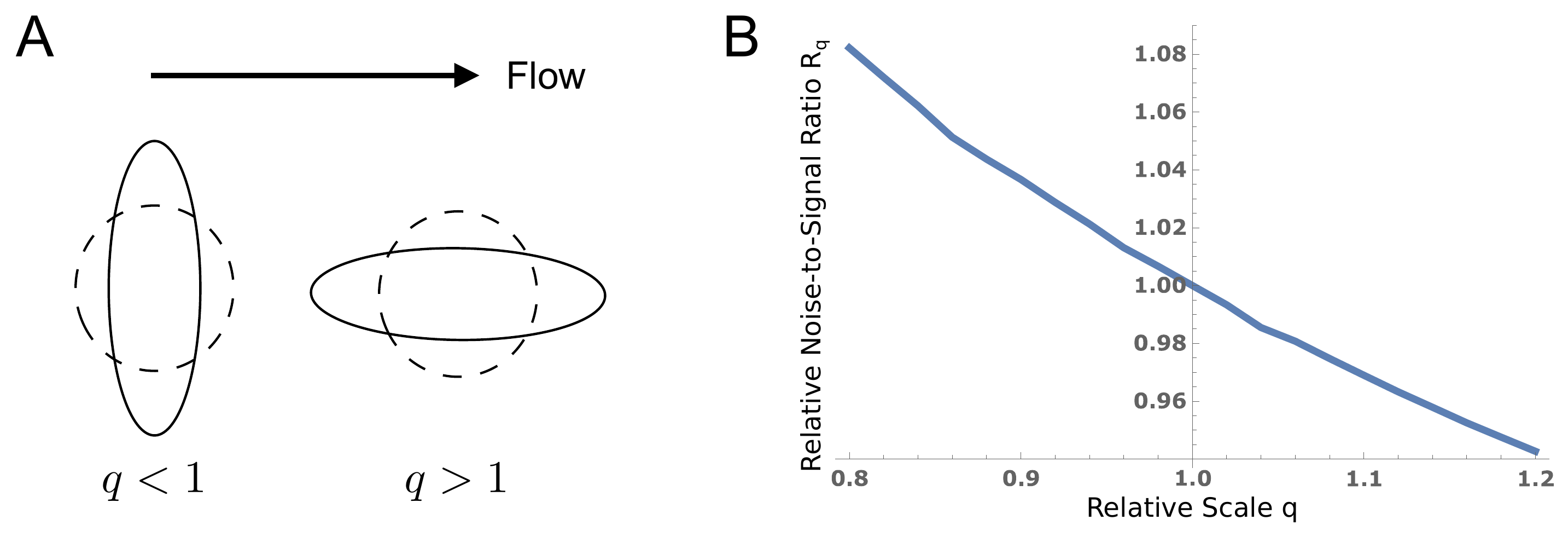}
	\end{center}
	\caption{Elongating in the flow direction can reduce sensory error. (A) We consider an ellipsoidal cell, where $q$ determines the elongation or compression while volume is conserved. (B) Sensory error relative to the spherical case (Eq.\ \ref{Rfinal}). Here $\tilde{\alpha}=0.74$, $\tilde{\beta} = 0.04$, $\epsilon=0.01$, and $r_{e,\max}/a = 100$.}
	\label{coord}
\end{figure}

The ellipsoidal coordinates are related to the cartesian ones via
\begin{equation}\label{elldef}
	\begin{gathered}
	x= r_eq^{-1/3}\sin \theta_e\cos \phi_e, \\
	y= r_eq^{-1/3}\sin \theta_e\sin \phi_e, \\
	z= r_eq^{2/3}\cos \theta_e, \\	
	\end{gathered}
\end{equation}
where the angular variables $(\theta_e,\phi_e)$ have the same ranges as the spherical angles $(\theta,\phi)$: $\theta_e\in(0,\pi)$ and $\phi_e\in(0,2\pi)$. Ellipsoids are surfaces of constant $r_e$. $q>0$ is a parameter characterizing the deformation: the $z$-axis is compressed for $q<1$, and the $z$-axis is stretched for $q>1$, as illustrated in \ref{coord}A. The ellipsoidal coordinates reduce to spherical coordinates when $q=1$.

More generally, the relationship between the ellipsoidal and spherical coordinates is obtained by comparing Eq.\ \ref{elldef} with the standard spherical-cartesian relations,
\begin{equation} \label{sphdef}
\begin{gathered}
x= r\sin\theta\cos\phi, \\
y= r\sin\theta\sin\phi, \\
z= r\cos\theta, \\
\end{gathered}
\end{equation}
as follows.
First, Eq.\ \ref{elldef} implies that $r_e = q^{1/3} \sqrt{x^2 + y^2 +q^{-2}z^2}$. Inserting Eq.\ \ref{sphdef} for $x$, $y$, and $z$ then yields $r_e = rq^{1/3} \sqrt{1+(q^{-2}-1)\cos^2\theta}$.
Second, Eq.\ \ref{elldef} implies $\cos \theta_e = z/r_eq^{2/3}$. Inserting Eq.\ \ref{sphdef} for $z$ and the previous result for $r_e$ gives $\cos \theta_e = q^{-1} \cos\theta/\sqrt{1 +(q^{-2}-1)\cos^2\theta}$.
Third, by considering the ratio $y/x$ in both Eq.\ \ref{elldef} and Eq.\ \ref{sphdef}, one sees that $\tan \phi_e=\tan \phi$, or $\phi_e = \phi$. In summary,
\begin{equation}\label{ell2sph}
\begin{gathered}
r_e = rq^{1/3} \sqrt{1+(q^{-2}-1)\cos^2\theta}, \\
\cos \theta_e = \frac{q^{-1} \cos\theta}{\sqrt{1 +(q^{-2}-1)\cos^2\theta}}, \\
\phi_e = \phi. \\
\end{gathered}
\end{equation}
Eq.\ \ref{ell2sph} gives the ellipsoidal coordinates in terms of spherical coordinates. The inverse is
\begin{equation}\label{sph2ell}
\begin{gathered}
r = r_e q^{-1/3}\sqrt{1+(q^2-1)\cos^2 \theta_e}, \\
\cos\theta = \frac{q \cos \theta_e}{\sqrt{1+(q^2-1)\cos^2 \theta_e}}, \\
\phi = \phi_e. \\
\end{gathered}
\end{equation}

The parametrization in Eq.\ \ref{elldef} explicitly conserves the volume of the ellipsoid. To see this fact, we recall that if the $x$, $y$, and $z$ semi-axis lengths are $A$, $B$, and $C$ respectively, then the volume of the ellipsoid is 
\begin{equation}
	V = \frac{4\pi}{3}ABC.
\end{equation}
Taking $A=B=r_eq^{-1/3}$ and $C=r_eq^{2/3}$, we see that the ellipsoid has the same volume as a sphere of radius $r_e>0$, independent of $q$. Note that $q$ is the ratio of axis lengths, as $q=C/A$.

Although the volume is conserved, the surface area is not.
This means that $\alpha$ and $\beta$ should change with $q$ in order to keep the total number of receptors on the surface and the rate of secretion constant, respectively. To account for this, we need the area element. A point on the surface of an ellipse in cartesian components is 
\begin{equation}\label{eq:param}
	\vec{r} =r_eq^{-1/3} \left\langle \sin \theta_e\cos \phi_e, \sin \theta_e\sin \phi_e, q \cos \theta_e \right\rangle.
\end{equation}
The area element may be computed as the cross-product
\begin{equation}
	dS_{r_e} = || \partial_{\theta_e} \vec{r} \times \partial_{\phi_e} \vec{r}||d\theta_e d\phi_e
	= r_e^2 q^{1/3} \sin \theta_e \sqrt{1+(q^{-2}-1)\cos^2 \theta_e}\ d\theta_e d\phi_e. 
\end{equation}
Letting $S_{r_e}$ denote the integral of $dS_{r_e}$ over the full ranges of the angular variables, we take
\begin{equation}\label{alphabeta}
	\alpha_{e} = \frac{4\pi a^2\alpha }{S_{a}}, \qquad \beta_{e} = \frac{4\pi a^2\beta }{S_{a}}.
\end{equation}
We perform the integral $S_{a}$ numerically.

\subsection{Flow lines}

Next we find the laminar flow lines around the ellipsoid, where the flow points along the stretched/compressed axis (Fig. \ref{coord}A). We do so following Ref.\ [46] of the main text, which specifies a numerical method for calculating flow lines in the laminar limit of the incompressible Navier-Stokes equations around an object with azimuthal symmetry.
We report the solution of Ref.\ [46] here in spherical coordinates, and then exploit our ellipsoidal coordinates when imposing the boundary conditions.

The flow velocity can be written
\begin{equation}\label{eq:axial}
	\vec{v} = \nabla\times\left(\frac{\psi(r,\theta)\hat{\phi}}{r\sin\theta}\right),
\end{equation}
or in terms of its components,
\begin{equation}\label{eq:velVecComp}
\vec{v}\cdot\hat{r} = \frac{\partial_{\theta}\psi}{r^2 \sin\theta}, \quad \vec{v}\cdot\hat{\theta} = -\frac{\partial_{r}\psi}{r \sin\theta}, \quad \vec{v}\cdot\hat{\phi}=0,
\end{equation}
where $\psi$ is the so-called stream function, and the last expression reflects the azimuthal symmetry.
The general solution for $\psi$ given in Ref.\ [46] is
\begin{equation}
\psi(r,\theta)=\sum\limits_{n=2}^{\infty} \left(a_n r^{-n+1} +b_n r^{-n+3} +c_n r^n +d_n r^{n+2} \right) C_{n}^{(-1/2)}(\cos\theta),
\end{equation}
where the $C_{n}^{(\mu)}$ are Gegenbauer polynomials.

We solve for the coefficients by imposing the boundary conditions. The flow at spatial infinity should point in the $\hat{z}$ direction
\begin{equation}
	\lim\limits_{r\rightarrow\infty} \vec{v} = v_0 \hat{z} = v_0(\cos\theta\hat{r} - \sin\theta\hat{\theta}).
\end{equation}
Considering Eq. \ref{eq:velVecComp}, we see that this holds if we have the asymptotic relation for large $r$
\begin{equation}
	\psi \sim \frac{v_0 r^2}{2}\sin^2\theta = \frac{v_0 r^2}{2}(1-\cos^2\theta).
\end{equation}
Aside from the factor $v_0 r^2$, the final expression is exactly $C^{(-1/2)}_2(\cos\theta)$. This can be used to solve for the $c_n$ and $d_n$ coefficients:
\begin{equation}
c_n = v_0 \delta_{n,2}, \quad d_n=0.
\end{equation}
The general solution now takes the form
\begin{equation} \label{psinow}
\psi(r,\theta)=  \frac{v_0 r^2}{2}(1-\cos^2\theta) + \sum\limits_{n=2}^{\infty} \left(a_n r^{-n+1} +b_n r^{-n+3}\right) C_{n}^{(-1/2)}(\cos\theta).
\end{equation}

The remaining coefficients are determined by requiring the fluid velocity to vanish at the surface of the cell, which implies
\begin{equation}\label{bcs}
	\frac{\partial \psi}{\partial r}=0, \qquad \frac{\partial \psi}{\partial (\cos\theta)}=0
\end{equation}
there. The relationship between $r$ and $\theta$ on the surface of the cell is given by the first line of Eq.\ \ref{ell2sph} with $r_e=a$,
\begin{equation}\label{rsurf}
r = \frac{a}{q^{1/3} \sqrt{1+(q^{-2}-1)\cos^2\theta}}.
\end{equation}
We use Eq.\ \ref{rsurf} to solve for the coefficients $a_n$ and $b_n$ in Eq.\ \ref{psinow} using the following sampling procedure from Ref.\ [46]. We sample $m$ points uniform randomly in $\cos\theta$ on the surface of the ellipsoid. For each point, we have two equations that result from inserting Eq.\ \ref{psinow} into the two boundary conditions (Eq.\ \ref{bcs}) with $r$ written in terms of $\cos\theta$ according to Eq.\ \ref{rsurf}. This gives $2m$ equations. We truncate the sum in Eq.\ \ref{psinow} at $n_{\max} =m+1$. This gives $2m$ unknowns (the coefficients $\{a_n\}_2^{m+1}$ and $\{b_n\}_2^{m+1}$).
The resulting linear system in the coefficients is solved by matrix inversion. Numerically, the matrix may be singular, and therefore we use the singular value decomposition. Once we solve for $a_n$ and $b_n$, the flow lines follow from Eq.\ \ref{eq:velVecComp}.

\subsection{Convection-diffusion equation}

Given the flow lines $\vec{v}$, we numerically solve the convection-diffusion equation (Eq.\ 2 of the main text),
\begin{equation}\label{dd2}
0 = D\nabla^2\bar{c} - \vec{v}\cdot\vec{\nabla}\bar{c}.
\end{equation}
This equation is subject to the secretion/absorption boundary condition at the ellipsoidal cell surface (analogous to Eq.\ 3 of the main text),
\begin{equation}\label{bc2}
-D\hat{n}\cdot\vec{\nabla} \bar{c}|_{r_e=a} = \beta_{e} -\alpha_{e}\bar{c}|_{r_e=a},
\end{equation}
where $\hat{n}$ is the outward-pointing unit vector orthogonal to ellipsoid, and $\alpha_e$ and $\beta_e$ are given in Eq.\ \ref{alphabeta}.
To solve Eq.\ \ref{dd2} subject to the ellipsoidal boundary condition, we derive the forms of the Laplacian $\nabla^2\bar{c}$ and convective term $\vec{v}\cdot\vec{\nabla}\bar{c}$ in ellipsoidal coordinates. Because the coordinates are non-orthogonal, it is most convenient to use the language of differential geometry (see Ref.\ [47] of the main text) to derive these forms.

\subsubsection{Laplacian in ellipsoidal coordinates}

Suppose that we change from one set of coordinates $\{x^{\mu}\}$ to another $\{y^{\mu'}\}$. A vector that transforms covariantly transforms like the chain rule for derivatives
\begin{equation}
	V'_{\mu'} = \frac{\partial x^{\mu}}{\partial y^{\mu'}} V_{\mu}.
\end{equation}
In the usual cartesian coordinates, we may write the derivative of a scalar in the direction of $\vec{v}$ as $\nabla_v = V^{\mu} \partial_{\mu}$, where the $V^{\mu}$ transform contravariantly
\begin{equation}
V'^{\mu'} = \frac{\partial y^{\mu'}}{\partial x^{\mu}} V^{\mu}
\end{equation}
under a change of coordinates. A tensor is an object with multiple indices, where each index transforms covariantly or contravariantly independently. The basis vectors on a manifold can change from point to point, so one must specify a curve to transport vectors and covectors along to define derivatives. This method of using transport to differentiate is called the covariant derivative, and its components for a general tensor take the form
\begin{equation}
	\nabla_{\mu} T^{\{\alpha\}}_{\{\beta\}} = \partial_{\mu} T^{\{\alpha\}}_{\{\beta\}} +\sum_{\alpha_{i}\in\{\alpha\}}\Gamma_{\rho\mu}^{\alpha_{i}}T^{\{\alpha|\alpha_{i}\to\rho\}}_{\{\beta\}}
	-\sum_{\beta_{i}\in\{\beta\}}\Gamma_{\beta_{i}\mu}^{\rho}T^{\{\alpha\}}_{\{\beta|\beta_{i}\to\rho\}},
\end{equation}
where $\{\alpha|\alpha_i\rightarrow \rho\}$ means that the $i$-th index has been changed to $\rho$ and summed over, $\partial_{\mu}$ is ordinary differentiation with respect to the coordinates, and the $\Gamma^{\rho}_{\mu\nu}$ are the Christoffel symbols. The Christoffel symbols are not tensorial and are defined in terms of derivatives of the metric tensor $g_{\mu\nu}$
\begin{equation}
\Gamma_{\mu\nu}^{\rho} = \frac{g^{\rho\alpha}}{2}\left(\partial_{\nu} g_{\mu\alpha}+\partial_{\mu} g_{\nu\alpha}-\partial_{\alpha} g_{\mu\nu}\right).
\end{equation}
The metric tensor $g_{\mu\nu}$ is defined in terms of the differential arc length $d\ell$ of curves
\begin{equation}
	(d\ell)^2 = g_{\mu\nu}dx^{\mu}dx^{\nu}.
\end{equation}
The metric tensor $g_{\mu\nu}$ may be used to lower indices, converting a contravariant vector to a covariant one. The inverse metric $g^{\mu\nu}$ may raise indices and perform the inverse conversion. When thought of as matrices, the inverse metric may be computed as the inverse of the metric.

We know that the metric in cartesian coordinates $g_{{\rm cart},\mu\nu}$ is the identity matrix. It is often easier to compute the matrix of partial derivatives $\partial x^{\mu}/\partial y^{\nu}$, where  $x^{\mu}=(x,y,z)$ denotes cartesian coordinates and $y^{\mu}$ denotes any new coordinates. We can use this to compute the metric
\begin{equation}
	g_{\text{new},\mu\nu} = \frac{\partial x^{\mu'}}{\partial y^{\mu}} \frac{\partial x^{\nu'}}{\partial y^{\nu}} g_{{\rm cart},\mu'\nu'}.
\end{equation}
Letting $y^{\mu}=(r_e,\theta_e,\phi_e)$ be the ellipsoidal coordinates, the metric tensor in ellipsoidal coordinates is
\begin{equation}
g_{e,\mu\nu} = \begin{bmatrix}
q^{\frac{4}{3}}\cos^2 \theta_e+q^{-\frac{2}{3}}\sin^2 \theta_e & r_e\sin \theta_e\cos \theta_e\left(q^{-\frac{2}{3}}-q^{\frac{4}{3}}\right) & 0 \\
r_e\sin \theta_e\cos \theta_e\left(q^{-\frac{2}{3}}-q^{\frac{4}{3}}\right) & r_e^{2}\left(q^{\frac{4}{3}}\sin^2 \theta_e+q^{-\frac{2}{3}}\cos^2 \theta_e\right) & 0 \\
0 & 0 & q^{-\frac{2}{3}}r_e^{2}\sin^2 \theta_e \end{bmatrix}.
\end{equation}
Inverting this gives the inverse metric
\begin{equation}
g_{e}^{\mu\nu} = \begin{bmatrix}
q^{\frac{2}{3}}\sin^2 \theta_e+q^{-\frac{4}{3}}\cos^2 \theta_e & \frac{1}{r_e}\sin \theta_e\cos \theta_e\left(q^{\frac{2}{3}}-q^{-\frac{4}{3}}\right) & 0 \\
\frac{1}{r_e}\sin \theta_e\cos \theta_e\left(q^{\frac{2}{3}}-q^{-\frac{4}{3}}\right) & \frac{1}{r_e^{2}}\left(q^{\frac{2}{3}}\cos^2 \theta_e+q^{-\frac{4}{3}}\sin^2 \theta_e\right) & 0 \\
0 & 0 & \frac{q^{\frac{2}{3}}}{r_e^{2}\sin^2 \theta_e} \end{bmatrix}.
\end{equation} 
With the metric and inverse metric, we may compute the Christoffel symbols
\begin{equation}
\begin{gathered}
\Gamma_{e,\mu\nu}^{1} = \begin{bmatrix}
0 & 0 & 0 \\
0 & -r_e & 0 \\
0 & 0 & -r_e\sin^2 \theta_e\end{bmatrix}, \quad
\Gamma_{e,\mu\nu}^{2} = \begin{bmatrix}
0 & \frac{1}{r_e} & 0 \\
\frac{1}{r_e} & 0 & 0 \\
0 & 0 & -\sin \theta_e\cos \theta_e\end{bmatrix}, \\
\Gamma_{e,\mu\nu}^{3} = \begin{bmatrix}
0 & 0 & \frac{1}{r_e} \\
0 & 0 & \frac{\cos \theta_e}{\sin \theta_e} \\
\frac{1}{r_e} & \frac{\cos \theta_e}{\sin \theta_e} & 0\end{bmatrix}.
\end{gathered}
\end{equation}
We can use this to find the gradient and Laplacian of a scalar in the new coordinates. The gradient describes the components of the covariant derivative, which is just the ordinary derivative, as scalars are invariant under changes of coordinates
\begin{equation}
	\nabla_{\mu} f=\partial_{\mu}f.
\end{equation}
The scalar Laplacian is the divergence of the gradient, which is
\begin{equation} \label{lapl}
\nabla^2 f = g_{e}^{\mu\nu} \nabla_{\mu} \nabla_{\nu}f =g_e^{\mu\nu}\left(\partial_{\mu}\partial_{\nu}f-\Gamma_{e,\mu\nu}^{\rho}\partial_{\rho} f\right).
\end{equation}
Using the expressions derived above, we find
\begin{align}
\nabla^{2}f &= q^{\frac{2}{3}}\left\lbrack\left(\sin^2 \theta_e+q^{-2}\cos^2 \theta_e\right)\frac{\partial^{2}f}{\partial r_e^{2}}+\frac{1}{r_e}\left(1+\cos^2 \theta_e+q^{-2}\sin^2 \theta_e\right)\frac{\partial f}{\partial r_e}\right. \nonumber\\
&\quad +\frac{1}{r_e^{2}}\left(\cos^2 \theta_e+q^{-2}\sin^2 \theta_e\right)\frac{\partial^{2}f}{\partial \theta_e^{2}}+\frac{1}{r_e^{2}}\left(\frac{\cos \theta_e}{\sin \theta_e}-2\left(1-q^{-2}\right)\sin \theta_e\cos \theta_e\right)\frac{\partial f}{\partial \theta_e} \nonumber\\
&\quad \left.+\frac{1}{r_e^{2}\sin^2 \theta_e}\frac{\partial^{2}f}{\partial \phi_e^{2}}+\frac{2}{r_e}\left(1-q^{-2}\right)\sin \theta_e\cos \theta_e\frac{\partial^{2}f}{\partial r_e\partial \theta_e}\right\rbrack.
\end{align}
This is the Laplacian in ellipsoidal coordinates.

\subsubsection{Convective term in ellipsoidal coordinates}

The convective term can be written in contravariant form as
\begin{equation}
\vec{v}\cdot \vec{\nabla} \bar{c} = V^{\mu}\partial_{\mu} \bar{c}.
\end{equation}
The gradient vector $\partial_\mu = (\partial_{r_e}, \partial_{\theta_e}, \partial_{\phi_e})$ is straightforward to write in ellipsoidal coordinates, but for the velocity vector it is easiest to transform to spherical coordinates,
\begin{equation} \label{convell}
	\vec{v}\cdot \vec{\nabla} \bar{c} = \frac{\partial y^{\mu}}{\partial x^{\nu}} V_{{\rm sph}}^{\nu}\partial_{\mu}\bar{c},
\end{equation}
and then write $\partial y^{\mu}/\partial x^{\nu}$ and $V_{{\rm sph}}^{\nu}$ in terms of $r_e$, $\theta_e$, and $\phi_e$. The former is obtained by differentiating Eq.\ \ref{ell2sph}, and then inserting Eq.\ \ref{sph2ell} into the results,
\begin{equation} \label{yx}
\frac{\partial y^{\mu}}{\partial x^{\nu}} = \frac{\partial(r_e,\theta_e,\phi_e)}{\partial(r,\theta,\phi)} = \begin{bmatrix}
\frac{q^{1/3}}{\sqrt{1+(q^{2}-1)\cos^2 \theta_e}} & r_eq^{-1}(q^{2}-1)\sin \theta_e\cos \theta_e &0 \\
0 & \frac{1}{2}q^{-1}\left[1+q^{2} +(q^{2}-1)\cos(2\theta_e)\right] & 0 \\
0 & 0 & 1
\end{bmatrix}.
\end{equation}
The latter is obtained by writing the curl (Eq.\ \ref{eq:axial}) in tensor notation,
\begin{equation} \label{Vtens}
V^{\alpha} = \epsilon^{\alpha\beta\gamma} \partial_{\beta} \Psi_{\gamma},
\end{equation}
where $\vec{\Psi} = \psi \hat{\phi}/r\sin\theta$ and $\epsilon^{\alpha\beta\gamma}$ is the Levi-Civita tensor. For orthogonal coordinate systems, like spherical coordinates, the Levi-Civita tensor may be written in terms of the Levi-Civita symbol $\tilde{\epsilon}$
\begin{equation}
\epsilon^{\alpha\beta\gamma} = \sqrt{\det(g^{\mu\nu})} \tilde{\epsilon}^{\alpha\beta\gamma},
\end{equation}
where $\tilde{\epsilon}$ is $+1$ for even permutations of $(1,2,3)$, $-1$ for odd permutations, and $0$ for repeated indices. The metric in spherical coordinates is well-known (and may be obtained from our ellipsoidal metric by taking $q\rightarrow1$),
\begin{equation}
g_{\text{sph}}^{\mu\nu} = \begin{bmatrix}
1 & 0 & 0 \\
0 & \dfrac{1}{r^{2}} & 0 \\
0 & 0 & \dfrac{1}{r^{2}\sin^{2}\theta}
\end{bmatrix},
\end{equation}
which implies that
\begin{equation}
\sqrt{\det(g_{\text{sph}}^{\mu\nu})} = \frac{1}{r^2\sin\theta}.
\end{equation}

However, the basis vectors that are commonly used in differential geometry are not normalized like $\hat{\phi}$. To find the basis vectors, we will start from the usual cartesian basis vectors, which are normalized and take the familiar form, and transform them. We expect $\hat{\phi}$ to be proportional to $\vec{e}_{\text{sph}}^{\text{ }3}$. We need the inverse Jacobian
\begin{equation}
\frac{\partial x^{\mu}}{\partial y^{\nu}} = \frac{\partial(x,y,z)}{\partial(r,\theta,\phi)} = 
\begin{bmatrix}
\sin\theta\cos\phi & r\cos\theta\cos\phi & -r\sin\theta\sin\phi \\
\sin\theta\sin\phi & r\cos\theta\sin\phi & r\sin\theta\cos\phi \\
\cos\theta & -r\sin\theta & 0
\end{bmatrix}.
\end{equation}
We compute $\vec{e}_{\text{sph}}^{\text{ } 3}$
\begin{equation}
\begin{aligned}
\vec{e}_{\text{sph}}^{\text{ } 3} &= g_{\text{sph}}^{3\mu} \frac{\partial x^{\nu}}{\partial y^{\mu}} \vec{e}_{\text{car},\nu} = g_{\text{sph}}^{33} \frac{\partial x^{\nu}}{\partial y^{3}} \vec{e}_{\text{car},\nu} = \frac{r\sin\theta(-\sin\phi \vec{e}_{\text{car},1}+\cos\phi \vec{e}_{\text{car},2})}{r^2 \sin^2\theta} , \\
&= \frac{-\sin\phi \vec{e}_{\text{car},1}+\cos\phi \vec{e}_{\text{car},2}}{r\sin\theta} =\frac{\hat{\phi}}{r\sin\theta}.
\end{aligned}
\end{equation}
We find the covariant components of $\vec{\Psi}$ in spherical coordinates by writing $\vec{\Psi} = \Psi_{\mu} \vec{e}_{\text{sph}}^{\text{ } \mu}$, which implies
\begin{equation}
	\Psi_1 = 0, \quad \Psi_2 = 0, \quad \Psi_3 =\psi.
\end{equation}
Using this to evaluate Eq.\ \ref{Vtens} gives
\begin{equation} \label{Vsph}
V_{\text{sph}}^{1} = \frac{\partial_{\theta}\psi}{r^2 \sin\theta}, \quad V_{\text{sph}}^{2} = - \dfrac{\partial_{r}\psi}{r^2 \sin\theta}, \quad V_{\text{sph}}^{3} =0.
\end{equation}
Note that these are not the same as the components of $\vec{v}$ along the unit vectors in spherical coordinates (Eq.\ \ref{eq:velVecComp}). They are slightly different because they are the components along the covariant basis vectors in spherical coordinates.

Thus, the convective term in ellipsoidal coordinates is given by Eq.\ \ref{convell}, with $\partial y^{\mu}/\partial x^{\nu}$ given by Eq.\ \ref{yx} and $V_{{\rm sph}}^{\nu}$ given by Eq.\ \ref{Vsph}. In Eq.\ \ref{Vsph}, $\psi$ is given by Eq.\ \ref{psinow}, and $r$ and $\theta$ are converted to ellipsoidal coordinates via Eq.\ \ref{sph2ell}.

\subsection{Relative error}

Finally, after obtaining the concentration $\bar{c}$ from Eq.\ \ref{dd2}, we calculate the mean and variance of the anisotropy measure. The anisotropy measure is defined analogously to Eq.\ 6 of the main text, as
\begin{equation}
\label{Aasupp}
A_e = \frac{1}{\bar{N}}\int_0^T dt \int dS_a \alpha_e c(a,\theta_e,\phi_e,t) \cos\theta,
\end{equation}
where
\begin{equation} \label{Nbaru}
\bar{N} = T \int dS_a \alpha_e \bar{c}(a,\theta_e)
\end{equation}
is the mean number of absorbed molecules in time $T$. The mean of Eq.\ \ref{Aasupp} is
\begin{equation} \label{Abaru}
\bar{A}_e = \frac{T}{\bar{N}} \int dS_a \alpha_e \bar{c}(a,\theta_e) \cos\theta.
\end{equation}
Eqs.\ \ref{Nbaru} and \ref{Abaru} are evaluated numerically using $\bar{c}$, where we write $\cos\theta$ in terms of $\cos \theta_e$ according to Eq.\ \ref{sph2ell}.

The variance of Eq.\ \ref{Aasupp} is $1/\bar{N}$, just as in Eq.\ 8 of the main text. To prove this fact, we use a generalization of the argument in Section V above. Specifically, Eq.\ \ref{eq:2mom} still holds for the statistics in the $\hat{z}$ direction, and Eq.\ \ref{Poissrand} still holds for the Poissonian $N$ in general. However, the variance of $A_e$ is no longer the variance in the $\hat{z}$ direction multiplied by a factor of three (Eq.\ \ref{var3}) because the ellipsoid breaks the spherical symmetry. Instead, we must write the components from the $\hat{x}$, $\hat{y}$, and $\hat{z}$ directions explicitly,
\begin{equation} \label{varAe}
\sigma_{A_e}^2 = \frac{1}{\bar{N}^2} \left[\text{Var}\left( \sum_{i=1}^N \sin\theta_i\cos\phi_i \right)
	+ \text{Var}\left( \sum_{i=1}^N \sin\theta_i\cos\phi_i \right)
	+ \text{Var}\left( \sum_{i=1}^N \cos\theta_i \right)\right].
\end{equation}
Nonetheless, we can still write the analogs of Eq.\ \ref{eq:2mom} explicitly for the $\hat{x}$, $\hat{y}$, and $\hat{z}$ directions,
\begin{align}
\label{2mom1}
\left\langle \left( \sum_{i=1}^N \sin\theta_i\cos\phi_i \right)^2 \right\rangle
	&= \left\langle \sum_{i=1}^N \sin^2\theta_i\cos^2\phi_i \right\rangle
	+ \left\langle \sum_{i\neq j} \sin\theta_i\cos\phi_i \sin\theta_j\cos\phi_j  \right\rangle \nonumber \\
	&= \braket{N} \braket{\sin^2\theta\cos^2\phi} +\braket{N(N-1)} \braket{\sin\theta\cos\phi}^2, \\
\left\langle \left( \sum_{i=1}^N \sin\theta_i\sin\phi_i \right)^2 \right\rangle
	&= \left\langle \sum_{i=1}^N \sin^2\theta_i\sin^2\phi_i \right\rangle
	+ \left\langle \sum_{i\neq j} \sin\theta_i\sin\phi_i \sin\theta_j\sin\phi_j  \right\rangle \nonumber \\
	&= \braket{N} \braket{\sin^2\theta\sin^2\phi} +\braket{N(N-1)} \braket{\sin\theta\sin\phi}^2, \\
\label{2mom3}
\left\langle \left( \sum_{i=1}^N \cos\theta_i \right)^2 \right\rangle
	&= \left\langle \sum_{i=1}^N \cos^2\theta_i \right\rangle
		+ \left\langle \sum_{i\neq j} \cos\theta_i \cos\theta_j  \right\rangle
	= \braket{N} \braket{\cos^2\theta} +\braket{N(N-1)} \braket{\cos\theta}^2.
\end{align}
Due to Eq.\ \ref{Poissrand}, the final terms in Eqs.\ \ref{2mom1}-\ref{2mom3} are still the squares of the means. Therefore, Eq.\ \ref{varAe} becomes
\begin{equation}
\sigma_{A_e}^2 = \frac{1}{\bar{N}^2} \left[\bar{N}\braket{\sin^2\theta\cos^2\phi}
	+ \bar{N}\braket{\sin^2\theta\sin^2\phi} + \bar{N}\braket{\cos^2\theta} \right]
	= \frac{1}{\bar{N}},
\end{equation}
as we sought to prove.

\subsection{Results}

For a given value of the ellipsoidal scale factor $q$, we compute the flow lines according to section B using $m=50$ points, and we solve the convection-diffusion equation for $\bar{c}$ according to section C. The ellipsoidal boundary condition in Eq.\ \ref{bc2} 
is implemented in Mathematica using the ``NeumannValue" function. We also use an ellipsoid at $r_e = r_{e,\max}$ for the outer boundary, where we impose $\bar{c}|_{r_e=r_{e,\max}} =0$.
We compute the error $\sigma_{A_e}/\bar{A}_e$, relative to the spherical case $\sigma_A/\bar{A}$, as
\begin{equation} \label{Rfinal}
R_q = \frac{\sigma_{A_e}/\bar{A}_e}{\sigma_A/\bar{A}},
\end{equation}
where $\sigma_A/\bar{A}$ is also computed numerically.

Fig.\ \ref{coord}B shows the ratio $R_q$ over the range where $q$ deviates from 1 by as much as 20\%. We see that elongating in the flow direction ($q > 1$) decreases the sensory error, whereas compressing in the flow direction ($q < 1$) increases the sensory error. Going beyond this range in $q$ requires prohibitively large computational runtime, as more than $m = 50$ terms are required in the flow lines for numerical accuracy, which significantly increases the runtime of the numerical routine for solving the convection-diffusion equation. Nonetheless, we can extrapolate out to $q = 2$, which corresponds to a cell that is twice as long as it is wide, for a rough idea of the effect. Treating the result in Fig.\ \ref{coord}B as a line (although it is slightly concave up) indicates that elongation to this extent would reduce the sensory error by about 30\%. Thus, we conclude that elongation in the flow direction can lead to a moderate improvement in the precision of flow sensing.

\end{document}